\documentclass[final,5p,times,twocolumn]{elsarticle}

\usepackage{amssymb}
\usepackage{mathtools}  
\usepackage{amsmath}
\usepackage{amssymb}
\usepackage{tabulary}
\usepackage{booktabs}
\usepackage{enumitem}   
\usepackage{multicol}
\usepackage{lscape}

\usepackage{multirow,url}

\usepackage[bookmarks,bookmarksnumbered]{hyperref}
\hypersetup{colorlinks = true,linkcolor = blue,anchorcolor =red,citecolor = blue,filecolor = red,urlcolor = red,
            pdfauthor=author}

\journal{Computer, Speech and Language}

\usepackage{color}
\usepackage{xcolor}

\newcommand*\dd{\color{black}}
\newcommand*\sw{\color{black}}
\newcommand*\kh{\color{black}}
\newcommand*\tp{\color{black}}
\newcommand*\nk{\color{black}}
\newcommand*\sn{\color{black}}

\newcommand*\revision{\color{black}}
\newcommand*\rrevision{\color{black}}

\newcommand{\RN}[1]{%
  \textup{\uppercase\expandafter{\romannumeral#1}}%
}

\DeclareMathOperator*{\argmax}{argmax}

\begin{document}

\begin{frontmatter}


\cortext[contrib]{Authors contributed equally}

\title{A Review of Speaker Diarization: Recent Advances with Deep Learning}

\author[usc]{Tae Jin Park\corref{contrib}}
\author[microsoft]{Naoyuki Kanda\corref{contrib}}
\author[microsoft]{Dimitrios Dimitriadis\corref{contrib}}
\author[asapp]{Kyu J. Han\corref{contrib}}
\author[jhu]{Shinji Watanabe\corref{contrib}}
\author[usc]{Shrikanth Narayanan}

\address[usc]{University of Southern California, Los Angeles, USA}
\address[microsoft]{Microsoft, Redmond, USA}
\address[asapp]{ASAPP, Mountain View, USA}
\address[jhu]{Johns Hopkins University, Baltimore, USA}

\begin{abstract}
{\tp Speaker diarization 
is a task to label audio or video recordings with classes that correspond to speaker identity, or in short, a task to identify ``who spoke when''.} {\kh In the early years, speaker diarization algorithms were developed for speech recognition on multispeaker audio recordings to enable speaker adaptive processing. These algorithms also gained their own value as a standalone application over time to provide speaker-specific metainformation for downstream tasks such as audio retrieval. More recently, with the emergence of deep learning technology, which has driven revolutionary changes in research and practices across speech application domains, rapid advancements have been made for speaker diarization. In this paper, we review not only the historical development of speaker diarization technology but also the recent advancements in neural speaker diarization approaches. Furthermore, we discuss how speaker diarization systems have been integrated with speech recognition applications and how the recent surge of deep learning is leading the way of jointly modeling these two components to be complementary to each other. By considering such exciting technical trends, we believe that this paper is a valuable contribution to the community to provide a survey work by consolidating the recent developments with neural methods and thus facilitating further progress toward a more efficient speaker diarization.}

\end{abstract}
\newpage



\end{frontmatter}

\section{Introduction}
\label{introduction}

\begin{figure*}[t]
  \centering
   \includegraphics[width=\textwidth]{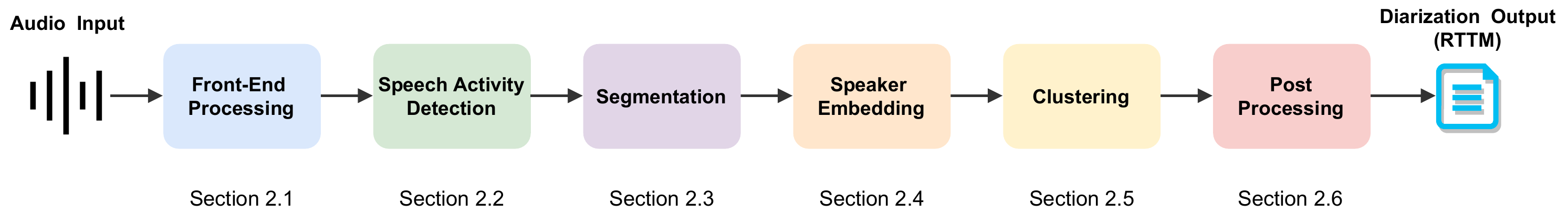}
  \caption{Traditional speaker diarization system.}
  \label{fig:modular}
\end{figure*}

{\kh ``Diarize'' means making a note or keeping an event in a diary. Speaker diarization, like keeping a record of events in such a diary, addresses the question of ``who spoke when''   \cite{tranter2003investigation,TrRe06,anguera2012speaker} by logging speaker-specific salient events on multiparticipant (or multispeaker) audio data. Throughout the diarization process, the audio data would be divided and clustered into groups of speech segments with the same speaker identity/label. As a result, salient events, such as non-speech/speech transition {\rrevision or} speaker turn {\rrevision changes}, are automatically {\rrevision detected}. In general, this process does not require any prior knowledge of the speakers, such as their real identity  or the number of participating speakers in the audio data. Thanks to its feature of separating audio streams by these speaker-specific events, speaker diarization can be effectively employed for indexing or analyzing  various types of audio data, e.g., audio/video broadcasts from media stations, conversations in conferences, personal videos from online social media or hand-held devices, court proceedings, business meetings, earnings reports in a financial sector, just to name a few.

Traditionally speaker diarization systems consist of multiple, independent sub-modules as presented in Fig.~\ref{fig:modular}. To mitigate any artifacts in acoustic environments, various front-end processing techniques, for example, speech enhancement, dereverberation, speech separation or target speaker extraction, are employed. Voice or speech activity detection (SAD) is then applied to separate speech from non-speech events. Raw speech signals in the selected speech portion {\rrevision are} transformed to acoustic features or embedding vectors. In the clustering stage, the {\revision transformed} speech portions are grouped and labeled by speaker classes and in the post-processing stage, the clustering results are further refined. Each of these sub-modules is optimized individually in general.} 

\subsection{Historical Development of Speaker Diarization} 
{\kh
During the early years of diarization technology (in the 1990s), the research objective was {\revision to benefit automatic speech recognition (ASR) on air traffic control dialogues and broadcast news recordings, by separating each speaker's speech segments and enabling speaker-adaptive training of acoustic models \cite{gish91segregation,siu1992,rohlicek92gist,Jain96recognitionof,padmanabhan96speaker,gauvain98partitioning,liu99fast}.} In this period some fundamental approaches {\rrevision for} {\revision measuring the distance between speech segments for} speaker change detection and clustering, such as generalized likelihood ratio (GLR) {\revision \cite{gish91segregation}} and Bayesian information criterion (BIC) {\revision \cite{Chen98speaker}}, were developed and quickly became the golden standard. All these efforts collectively laid out path{\revision s} to consolidate activities across research groups worldwide, leading to several research consortia and challenges in the early 2000s, among which there were the Augmented Multiparty Interaction (AMI) Consortium \cite{amiwebsite} supported by the European Commission and the RT Evaluation \cite{nistrtwebsite} hosted by the National Institute of Standards and Technology (NIST). These organizations, spanning over from a few years to a decade, {\rrevision fostered} further advancements on speaker diarization technologies across different data domains from broadcast news \cite{ajmera03asru,tranter04bn,reynolds05approaches,zhu05bic,meignier06csl} and conversational telephone speech (CTS) \cite{rosenberg02cts,liu03cts,tranter04cts,kenny10cts} to meeting conversations \cite{ajmera04meeting,jin04meeting,anguera06icassp,leeuwen07clear,vijayasenan09taslp}. The new approaches resulting from these advancements include, but not limited to, beamforming \cite{anguera07taslp}, information bottleneck clustering (IBC) \cite{vijayasenan09taslp}, variational Bayesian (VB) approaches \cite{valente10vb}, joint factor analysis (JFA) \cite{kenny10cts}.

{\revision Speaker specific representation in a total variability space derived from simplified JFA, known as i-vector \cite{Dehak+_2011}, {\rrevision found} great success in speaker recognition and was quickly adopted by speaker diarization systems as feature representation for short speech segments, segmented in an unsupervised fashion. i-{\rrevision V}ector successfully replaced its predecessors such as merely mel-frequency cepstral coefficient (MFCC) or speaker factors (or eigenvoices) \cite{eigenvoice} to bolster clustering performance in speaker diarization, being combined with principal component analysis (PCA) \cite{shum2011exploiting,shum2012use}, variational Bayesian Gaussian mixture model (VB-GMM) \cite{shum2013unsupervised}, mean shift \cite{senoussaoui2013study} and probabilistic linear discriminant analysis (PLDA) \cite{sell2014speaker}.  
}

Since the advent of deep learning in the 2010s, there has been a considerable amount of research to take advantage of powerful modeling capabilities of the neural networks for speaker diarization. One representative example is the extraction of the speaker embeddings  using neural networks, {\dd such as the d-vectors~\cite{Variani+_2014,HMBS2016,wang2018speaker} or the x-vectors~\cite{snyder2018x}, which most often are embedding vector representations based on the bottleneck layer output of a deep neural network (DNN) trained for speaker recognition. The shift from i-vector to these neural embeddings contributed to enhanced performance, easier training with more data~\cite{zhang2019fully}, and robustness against speaker variability and acoustic conditions.} {\kh More recently, end-to-end neural diarization (EEND) where individual sub-modules in the traditional speaker diarization systems (c.f., Fig.~\ref{fig:modular}) can be replaced by one neural network gets more attention with promising results \cite{fujita2019end,fujita2019end2}. This research direction, although not fully matured yet, could open up unprecedented opportunities to address challenges in the field of speaker diarization, such as, the joint optimization with other speech applications, with overlapping speech, if large-scale data is available for  training such powerful {\rrevision neural} network-based models.}
}

\subsection{Motivation} 
{\kh Till now, there are two well-rounded overview papers in the area of speaker diarization that survey the development of speaker diarization technology with different focuses. In \cite{TrRe06}, various speaker diarization systems and their subtasks in the context of broadcast news and CTS data are reviewed up to till mid 2000s. Thus, the historical progress of speaker diarization technology development in the 1990s and early 2000s are covered. Contrarily,  the focus of \cite{anguera2012speaker} was put more on speaker diarization for meeting speech and its respective challenges. This paper thus weighs more in the corresponding technologies to mitigate problems from the perspective of meeting environments, where there are usually more participants than broadcast news or CTS data and  multi-modal data is frequently available. Since these two papers were published, speaker diarization systems have gone through a lot of notable changes,  especially from the leap-frog advancements in deep learning approaches addressing technical challenges across multiple machine learning domains. We believe that this survey work is a valuable contribution to the community to consolidate the recent developments with neural methods and thus facilitate further progress toward a more efficient diarization.}

\subsection{Overview and Taxonomy of Speaker Diarization} 

\begin{table*}[t]\caption{Table of Taxonomy}
\label{tab:TableOfTaxonomy}
\begin{center}
{
\begin{tabular}{r|c|c}
\toprule
\multirow{2}{8em}{}& {\revision \textbf{Trained based on}} & {\revision\textbf{Trained based on}}\\
 &\textbf{Non-diarization Objective} & \textbf{Diarization Objective}\\
\hline
\multirow{2}{8em}{\textbf{\\Single-module Optimization}}  & \multirow{2}{14em}{\underline{\bf Section {\revision 2.1--2.6}}\\Front-end \cite{hershey2016deep,kolbaek2017multitalker,luo2019conv}, speaker embedding \cite{variani2014deep,snyder2017deep,snyder2018x}, SAD \cite{drugman2015voice}, etc. }  & \multirow{2}{14em}{\underline{\bf Section 3.1}\\Affinity matrix refinement \cite{wang2020speaker}, IDEC \cite{Dimitriadis19}, TS-VAD \cite{medennikov2020target}, etc.}\\
& \\
& \\
& \\
\hline
\multirow{6}{8em}{\textbf{\\Joint Optimization}} & \multirow{2}{14em}{{\revision\underline{\bf Section 2.7}}\\ {\revision VB-HMM \cite{diez2018speaker}, VBx \cite{diez2019bayesian}}\\\underline{\bf Out of scope}\\Joint front-end \& ASR \cite{yu2017recognizing,seki2018purely,chang2019end,kanda2019acoustic,kanda2019auxiliary,wang2020exploring}, joint speaker identification \& speech separation \cite{wang2019speech, han2020continuous}, etc.} & \multirow{2}{14em}{\underline{\bf Section 3.2}\\UIS-RNN~\cite{zhang2019fully}, RPN \cite{huang2020speaker}, online RSAN \cite{von2019all}, EEND \cite{fujita2019end,fujita2019end2}, etc. \\ 
 \underline{\bf Section 4}\\Joint ASR \& speaker diarization. \cite{Shafey2019,mao2020speech,kanda2019simultaneous,kanda2021investigation}, etc.} \\
 & \\
 & \\
 & \\
 & \\
 & \\
 & \\
\bottomrule
\end{tabular}
}
\end{center}
\end{table*}

{\kh Attempting to categorize the existing, most-diverse speaker diarization technologies, both in the space of  modularized speaker diarization systems before the deep learning era and those based on neural networks of recent years, a proper grouping would be helpful.}
{\nk The main categorization we adopt in this paper is based on two criteria,
 resulting total of four categories, as shown in Table \ref{tab:TableOfTaxonomy}.
The first criterion is whether the model is trained based on speaker diarization-oriented objective function or not.
Any trainable approaches to optimize models in a multispeaker situation and learn relations between speakers are categorized into the ``diarization objective'' class.
The second criterion is whether multiple modules
are jointly optimized toward some
objective function.
If a single sub-module is replaced into a trainable one, such
method is categorized into the ``Single-module optimization'' class.
Conversely, joint modeling of segmentation and clustering \cite{zhang2019fully}, joint modeling of
speech separation and speaker diarization \cite{von2019all} or fully end-to-end neural diarization system \cite{fujita2019end,fujita2019end2}
is categorized into the ``Joint optimization'' class.

Note that 
our intention of this categorization is to help readers
to quickly overview the broad development in the field,
and it is not our intention to divide the categories into superior-inferior.
Also, while we are aware of many techniques that fall into the category ``Non-Diarization Objective'' and ``Joint Optimization'' (e.g., joint front-end and ASR \cite{yu2017recognizing,seki2018purely,chang2019end,kanda2019acoustic,kanda2019auxiliary,wang2020exploring} and joint speaker identification and speech separation \cite{wang2019speech,han2020continuous}), 
we exclude them in the paper to focus on the review of speaker diarization techniques.
}

{\revision
\subsection{Diarization Evaluation Metrics}
}

\label{sec:Diarization Evaluation Metrics}
\subsubsection{Diarization Error Rate}
\label{subsubsec:DER}
{\tp The accuracy of speaker diarization system is measured using diarization error rate (DER) \cite{fiscus2006rich} where DER is the sum of three different error types: False alarm (FA) of speech, missed detection of speech and confusion between speaker labels. 
\begin{equation}
\label{eq:der}
    \textbf{DER} = \frac{ \text{FA} + \text{Missed} + \text{Speaker-Confusion}}{\textit{Total Duration of Time}}.
\end{equation}
To establish a one-to-one mapping between the hypothesis outputs and the reference transcript, Hungarian algorithm \cite{nist_hungarian} is employed. In the 2006 RT evaluation \cite{fiscus2006rich}, 0.25 s of ``no score" collar (also referred to as ``score collar'') is set around every boundary of reference segment to mitigate the effect of inconsistent notation and human errors in reference transcript and this evaluation scheme has been most widely used in speaker diarization studies.}
\subsubsection{Jaccard Error Rate}
\label{JER}
{ \tp 
The Jaccard error rate (JER) was first introduced in the DIHARD II evaluation. The goal of JER is to evaluate each speaker with equal weight. {\rrevision Unlike DER which is estimated for the whole utterance altogether, per-speaker error rates are first computed and then averaged to compute JER.
Specifically, JER is computed as follows.} 
\begin{equation}
\label{eq:jer}
\textbf{JER}=\frac{1}{N} \sum_{i}^{N_{ref}} \mathrm{\frac{\mathrm{FA_{i}}+\mathrm{MISS_{i}}}{\mathrm{TOTAL_{i}}}}.
\end{equation}
In Eq. (\ref{eq:jer}), $\mathrm{TOTAL}_i$ is the union of the $i$-th speaker's speaking time in the reference transcript and the $i$-th speaker's speaking time in the hypotheses. {\rrevision $N_{ref}$ is the number of speakers in the reference script. Note that the Speaker-Confusion in DER is reflected in the part of $\mathrm{FA_i}$ in the calculation of JER. Since JER is using union operation between reference and the hypotheses, JER never exceeds 100\%, whereas DER can become much larger than 100\%. DER and JER are highly correlated but if a subset of speakers are dominant in the given audio recording, JER tends to be higher than the ordinary case.}
}

\subsubsection{Word-level Diarization Error Rate}

{\tp
While DER is based on the duration of the speaking time of each speaker, word-level DER (WDER) is designed to measure the error that is caused in the lexical (output transcription) side. The motivation of WDER is the discrepancy between DER and the accuracy of the final transcript output since DER relies on the duration of the speaking time that is not always aligned with the word boundaries. The concept of word-breakage ratio was proposed in \citet{silovsky2012incorporation} where word-breakage shares similar idea with WDER. Unlike WDER, word-breakage ratio measures the number of speaker-change points occur inside a word boundary. The work in \citet{park2018multimodal} suggested the term WDER, evaluating the diarization output with ground-truth transcription. More recently, the joint ASR and speaker diarization system was evaluated in the WDER format in \citet{Shafey2019}. Although the way of calculating WDER would differ over the studies, but the underlying mechanism is that the diarization error is calculated by counting the correctly or incorrectly labeled words.
}

\subsection{Paper Organization}
{\kh The rest of the paper is organized as follows. 
\begin{itemize}
    \item In Section 2, we overview techniques belonging to the {\revision ``Non-diarization objective"} class in the proposed taxonomy, mostly those used in the traditional, modular speaker diarization systems. While there are some overlaps with the counterpart sections of the aforementioned two survey papers \cite{TrRe06,anguera2012speaker} in terms of reviewing notable developments in the past, this section would add more latest schemes in the corresponding components of the speaker diarization systems.
    \item  In Section 3, we discuss advancements mostly leveraging DNNs trained with the diarization objective where single sub-modules are independently optimized (Subsection 3.1) or jointly optimized (Subsection 3.2) toward fully end-to-end speaker diarization. 
    \item In Section 4, we present a perspective of how speaker diarization has been investigated in the context of ASR, reviewing historical interactions between these two domains to peek into the past, present and future of speaker diarization applications.
    \item Section 5 provides information on speaker diarization challenges and corpora to facilitate research activities and anchor technology advances. We also discuss evaluation metrics such as DER, JER and Word-level DER (WDER) in this section. 
    \item We share a few examples of how speaker diarization systems are employed in both research and industry practices in Section 6 and conclude this work in Section 7, providing summary and future challenges in speaker diarization.
\end{itemize}
}

\section{Modular Speaker Diarization Systems}
\label{Conventional Speaker Diarization Systems}
This section provides an overview of algorithms for speaker diarization belonging to the ``Non-diarization objective'' class, as shown in Table~\ref{tab:TableOfTaxonomy}. Each subsection in this section corresponds to the explanation of each module in the traditional speaker diarization system, as shown in Figure~\ref{fig:modular}.
In addition to the introductory explanation of each module, this section also summarizes the recent techniques within the module.

\subsection{Front-end Processing} 

{\sw
This section describes mostly front-end techniques, used for speech enhancement, dereverberation, speech separation, and speech extraction as part of the speaker diarization pipeline.
Let $s _{i, f, t} \in \mathbb{C}$ be the short-time Fourier Transform (STFT) representation of the source speaker $i$ on the frequency bin $f$ at frame $t$.
The observed noisy signal $x _{t, f}$ can be represented by a mixture of the source signals, a 
room impulse response $h_{i, f, t} \in \mathbb{C}$, and additive noise $n _{t, f} \in \mathbb{C}$,
\begin{equation}
    \label{eq:signal_model_stft}
    x _{t, f} = \sum _{i=1} ^K \sum _{\tau} h_{i, f, \tau} s _{i, f, t - \tau} + n _{t, f},
\end{equation}
where $K$ denotes the number of speakers present in the audio signal.

{\revision 
The aim of the front-end techniques described in this section is to estimate the original source signal} $\hat{\mathbf{x}}_{i, t}$ given the observation $\mathbf{X}=(\{x _{t, f}\}_{f})_t$ for the downstream diarization task,
\begin{equation}
    \hat{\mathbf{x}}_{i, t} = \mathrm{FrontEnd}(\mathbf{X}), \quad i = 1, \dots, K,
\end{equation}
where $\hat{\mathbf{x}}_{i, t} \in \mathbb{C} ^D$ denotes the $i$-th  speaker's estimated STFT spectrum with $D$ frequency bins at frame $t$.

Although there are numerous speech enhancement, {\revision dereverberation}, and separation algorithms, e.g., ~\cite{haeb2019speech,vincent2018audio,wang2018supervised}, herein most of the recent techniques used in the DIHARD challenge series \cite{sell2018diarization,ryant2019second,diez2018but}, LibriCSS meeting recognition task \cite{chen2020continuous,raj2020integration}, and the CHiME-6 challenge track 2 \cite{watanabe2020chime,arora2020jhu,medennikov2020stc} are covered.

\subsubsection{Speech Enhancement and Denoising}
\label{subsec:denoising}
{\revision
Speech enhancement techniques focus mainly on the suppression of the noise component of the noisy speech, which has shown a significant improvement thanks to deep learning.}
For example, long short-term memory (LSTM)-based speech enhancement \cite{gao2018densely,erdogan2015phase} is used as a front-end technique in the DIHARD II baseline \cite{ryant2019second}, i.e., 
\begin{equation}
    \hat{\mathbf{x}} _t = \mathrm{LSTM}(\mathbf{X}),
\end{equation}
where we only consider the single source example (i.e., $K = 1$) and omit the source index $i$.
This is a regression-based approach by minimizing the objective function,
\begin{equation}
    \mathcal{L} _{\mathrm{MSE}} = ||\mathbf{s} _{t} - \hat{\mathbf{x}} _t||^2.
\end{equation}
The log power spectrum or ideal ratio mask is often used as the target domain of the output $\mathbf{s} _{t}$.
Also, the speech enhancement used in  \cite{loizou2013speech} applies this objective function in each layer based on a progressive manner.

The effectiveness of the speech enhancement techniques can be enhanced via multichannel processing, including minimum variance distortionless response (MVDR) beamforming \cite{haeb2019speech}. \cite{raj2020integration} demonstrates the significant improvement of the DER from 18.3\% to 13.9\% in the LibriCSS meeting task based on mask-based MVDR beamforming \cite{heymann2016neural,erdogan2016improved}.

\subsubsection{Dereverberation} 
\label{subsec:dereverb}
Compared with other front-end techniques, the major dereverberation techniques used in various tasks are based on statistical signal processing methods.
One of the most widely used techniques is the weighted prediction error (WPE) based dereverberation \cite{nakatani2010speech,yoshioka2012generalization,drude2018nara}.

The basic idea of WPE, for the case of single source, i.e., $K = 1$, without  noise, is to decompose the original signal model Eq.~\eqref{eq:signal_model_stft} into the early reflection $x _{t, f} ^{\text{early}}$ and late reverberation $x _{t, f} ^{\text{late}}$ as follows:
\begin{align}
    x _{t, f} & = \sum _{\tau} h_{f, \tau} s _{f, t - \tau} = x _{t, f} ^{\text{early}} + x _{t, f} ^{\text{late}}.
\end{align}
WPE tries to estimate filter coefficients $\hat{h} ^{\text{wpe}} _{f, t} \in \mathbb{C}$, which maintain the early reflection while suppressing the late reverberation based on the maximum likelihood estimation.
\begin{equation}
    \hat{x} ^{\text{early}}  _{t, ,f} = x _{t, f} - \sum _{\tau = \Delta} ^L \hat{h} ^{\text{wpe}} _{f, \tau} x _{f, t - \tau},
\end{equation}
where $\Delta$ denotes the number of frames to split the early reflection and late reverberation, and $L$ denotes the filter size.

WPE is widely used as one of the gold standard front-end processing methods, e.g., it is part of the DIHARD and CHiME, which are both the baseline and the top-performing systems \cite{sell2018diarization,ryant2019second,diez2018but,watanabe2020chime,arora2020jhu}.
Although the performance improvement of WPE-based dereverberation is not  significant, it provides solid performance improvement across almost all tasks.
Moreover, WPE is based on linear filtering and since it does not introduce signal distortions, it can be  safely combined with downstream front-end and back-end processing steps.
Similar to the speech enhancement techniques, WPE-based {\revision dereverberation} demonstrates additional performance improvements when applied on multichannel signals.

\subsubsection{Speech Separation}
\label{subsec:ss-tse}
Speech separation is a promising family of techniques when the  overlapping speech regions are significant.
The effectiveness of multichannel speech separation based on beamforming
has been widely confirmed \cite{anguera07taslp,yoshioka2018recognizing,boeddecker2018front}. 
For example, in the CHiME-6 challenge \cite{watanabe2020chime},
guided source separation (GSS) \cite{boeddecker2018front} based multichannel speech extraction techniques have been used to achieve the top result.
On the other hand,
single-channel speech separation techniques \cite{hershey2016deep,kolbaek2017multitalker,luo2019conv} do not often  show any significant effectiveness in realistic multispeaker scenarios, such as the LibriCSS \cite{chen2020continuous} or the CHiME-6 tasks \cite{watanabe2020chime}, where speech signals are continuous and contain both overlapping and overlap-free speech regions. 
The single-channel speech separation systems often produce a redundant non-speech  
or even a duplicated speech signal for the non-overlap regions, and 
as such the ``leakage" of audio causes many false alarms of speech activity.
A leakage filtering method was proposed in \cite{xiao2020microsoft} tackle the problem, where  a significant improvement in the diarization performance was observed after including this processing step in  the top-ranked system on the VoxCeleb Speaker Recognition Challenge 2020 \cite{nagrani2020voxsrc}.
}

\subsection{Speech Activity Detection} 
{\revision 
SAD, also known as voice activity detection (VAD), distinguishes speech from non-speech such as background noise. SAD plays a significant role not only in speaker diarization but also in speaker recognition and speech recognition systems since SAD is a pre-processing step that can create errors that propagate through the whole pipeline.
A SAD system consists mostly of two major parts. The first one is a feature extraction front-end, where acoustic features such as {\rrevision 
 zero crossing rate \cite{itu1996silence},
pitch \cite{chengalvarayan1999robust},
signal energy \cite{woo2000robust}, 
higher order statistics in the linear predictive coding residual domain \cite{nemer2001robust}
or 
MFCCs are often used. The other part is a classifier, where a model predicts and decides whether the input-frame contains speech or not.
A system based on statistical models on spectrum \cite{sohn1999statistical}, Gaussian mixture models (GMMs) \cite{ng2012developing} and on Hidden Markov Models (HMMs) \cite{pfau2001multispeaker, sarikaya1998robust} has been traditionally used. 
After the deep learning approaches gained popularity in the speech signal processing field, numerous DNN-based systems, such as the ones based on MLP \cite{ryant2013speech}, convolutional neural network (CNN) \cite{thomas2014analyzing}, LSTM \cite{gelly2017optimization}, have been also proposed with superior performance to the traditional methods.}

}
The performance of SAD largely affects the overall performance of the speaker diarization system 
as it can create a significant amount of false positive salient events or miss speech segments~\cite{Haws+16}.
A common practice in speaker diarization tasks is to report DER with ``oracle SAD'' setup which indicates that the system output is using SAD output that is identical to the ground truth. 
Conversely, the system output with an actual speech activity detector is referred to as ``system SAD'' output.

\subsection{Segmentation}
\label{subsec:segmentation}
{\revision
In the context of speaker diarization, speech segmentation is a process of breaking the input audio stream into multiple segments to obtain speaker-uniform segments. Therefore, the output unit of the speaker diarization system is determined via a segmentation process. In general, speech segmentation methods for speaker diarization are divided into two major categories: Segmentation by speaker-change point detection and uniform segmentation.

Segmentation by the speaker-change point detection was the gold standard of the earlier speaker diarization systems, where speaker-change points are detected by comparing two hypotheses: $H_0$ assumes that both the left and right speech windows are from the same speaker, whereas $H_1$ assumes that the two speech windows are from the different speakers. 
{\rrevision 
To test these two hypotheses, metric-based approaches \cite{chen1998speaker, kemp2000strategies} were most widely applied. In metric-based
approaches, 
the distribution of the speech feature is assumed to follow a Gaussian distribution $\mathcal{N}(\mu, \Sigma)$ with mean $\mu$ and covariance $\Sigma$.
The two hypotheses $H_0$ and $H_1$ can be then represented as follows:
\begin{equation}
\begin{aligned}
\label{eq:two_hypotheses}
H_{0}: \;& \mathbf{x}_{1} \cdots \mathbf{x}_{N} \sim \mathcal{N}(\mu, \Sigma), \\
H_{1}: \;& \mathbf{x}_{1} \cdots \mathbf{x}_{i}  \sim \mathcal{N}\left(\mu_{1}, \Sigma_{1}\right), \\
       \;& \mathbf{x}_{i+1} \cdots \mathbf{x}_{N} \sim \mathcal{N}\left(\mu_{2}, \Sigma_{2}\right),
\end{aligned}
\end{equation}
where $(\mathbf{x}_{i}|i=1,\cdots,N)$ is a sequence of speech features in the interest of the hypothesis testing.
A slew of criteria for the metric-based approach were proposed to quantify the likelihood of the two hypotheses. The examples include the Kullback Leibler (KL) distance~\cite{siegler1997automatic}, Generalized Likelihood Ratio (GLR)~\cite{gish1991segregation, bonastre2000speaker, gangadharaiah2004novel} and BIC~\cite{chen1998speaker, tritschler1999improved}. Among these criteria, BIC has been the most widely used method followed by numerous variants \cite{delacourt2000distbic, mori2001speaker, ajmera2004robust, malegaonkar2006unsupervised}. Thus, in this section, we introduce BIC as a representative example for a metric-based method. If we apply BIC and to the hypotheses described in Eq. (\ref{eq:two_hypotheses}), a BIC value between two models from two hypotheses is expressed as follows:}
\begin{align}
\label{eq:bic_seg}
BIC(i)=N \log |\Sigma|-N_{1} \log \left|\Sigma_{1}\right|-N_{2} \log \left|\Sigma_{2}\right| -\lambda P,
\end{align}
where the sample covariance $\Sigma$ is from $\left\{\mathbf{x}_{1}, \cdots, \mathbf{x}_{N}\right\}$, $\Sigma_{1}$ is from $\left\{\mathbf{x}_{1}, \cdots, \mathbf{x}_{i}\right\}$ and $\Sigma_{2}$ is from $\left\{\mathbf{x}_{i+1}, \cdots, \mathbf{x}_{N}\right\}$ and $P$ is the penalty term \cite{chen1998speaker} defined as
\begin{align}
\label{eq:penalty_term}
   P = \frac{1}{2}\left(d+\frac{1}{2} d(d+1)\right) \log N,
\end{align}
where $d$ denotes the dimension of the feature; $N_1$ and $N_2$ are frame lengths of each window, respectively and $N = N_{1} + N_{2}$. The penalty weight $\lambda$ is generally set to $\lambda=1$. The change point is set when the following equation becomes true: 
\begin{align}
\label{eq:BIC_testing}
\left\{\max _{i} B I C(i)\right\}>0.
\end{align}
In general, if speech segmentation is done using the speaker-change point detection method, the length of each segment is not consistent. Therefore, after the advent of the i-vector~\cite{Dehak+_2011} and DNN-based embeddings \cite{variani2014deep, snyder2018x}, the segmentation based on speaker-change point detection was mostly replaced with uniform segmentation \cite{senoussaoui2013study, Sell18, wang2018speaker}, since varying lengths of the segment created an additional variability into the speaker representation and deteriorated the fidelity of the speaker representations. 

In uniform segmentation schemes, the given audio stream input is segmented with a fixed window length and overlap length. Thus, the unit duration of the speaker diarization output stays constant. However, the process of uniform segmentation of the input signals for diarization  poses some potential problems because it introduces a trade-off error related to the segment length. The segments created from the uniform segmentation need to be sufficiently short to safely assume that they do not contain multiple speakers. However, at the same time it is important to capture sufficient acoustic information to extract reliable speaker representations. 
}
\label{subsec:vad}

\subsection{Speaker Representations and {\rrevision Similarity Measure}} 

{\rrevision Speaker representation plays a crucial role for speaker diarization systems to measure the similarity between speech segments. This section will cover such speaker representation and also the similarity measure because they are tightly connected.} 
We first introduce metric-based approaches for similarity measures which were popular from the late 1990s to early 2000s {\rrevision in Section \ref{sec:GMM speaker model}}. 
We then introduce widely used speaker representations for speaker diarization systems that are usually employed together with the uniform segmentation method described in Section \ref{sec:Joint Factor Analysis and i-vector} and Section  \ref{sec:Neural Network Based Speaker Representations}. 

\label{subsec:embedding}

\subsubsection{Metric Based Similarity Measure}
\label{sec:GMM speaker model}
{\rrevision 
From the late 1990s to early 2000s, metric-based approaches were most commonly used for the similarity measurement between speech segments for speaker diarization systems. 
Methods used for speaker segmentation were also applied to measure the similarity between segments, such as KL distance~\cite{siegler1997automatic}, GLR~\cite{gish1991segregation, bonastre2000speaker, gangadharaiah2004novel}, and BIC~\cite{chen1998speaker, tritschler1999improved}. As with the case of the segmentation, the BIC-based method,
where the similarity between two segments are computed by Eq.~\eqref{eq:bic_seg}, was one of the most extensively used metrics due to its effectiveness and ease of implementation. 
Metric-based approaches are usually employed together with the segmentation approaches based on a speaker-change point detection.
The agglomerative hierarchical clustering (AHC) is often applied to obtain the diarization result, which will be detailed in Section \ref{subsubsec:AHC}.}



\subsubsection{Joint Factor Analysis, i-vector and PLDA}
\label{sec:Joint Factor Analysis and i-vector}

Before the advent of speaker representations such as i-vector~\cite{Dehak+_2011} or x-vector~\cite{snyder2018x}, {\revision Gaussian Mixture Model{\rrevision-based} Universal Background Model (GMM-UBM)~\cite{reynolds2000speaker} applied to acoustic features demonstrated success in speaker verification tasks. 
A UBM {\rrevision consists of} a large GMM (typically with 512 to 2048 mixtures) trained to represent the speaker-independent distribution of acoustic features. Thus, a GMM-UBM model can be described with the following quantities: mixture weights, mean values and covariance matrix of the mixtures. {\rrevision The log-likelihood ratio between a speaker-adapted GMM and the speaker-independent GMM-UBM is used for speaker verification.} Despite the success on modeling the speaker identity, GMM-UBM based speaker verification systems have suffered from intersession variability \cite{kenny2007speaker}, which is the variability exhibited by a given speaker from one recording session to another. Such difficulty occurs because the relevance maximum a posteriori (MAP) adaptation step during the speaker enrollment process in the GMM-UBM based speaker verification systems not only captures the speaker-specific characteristics of the speech, but also unwanted channel noise and other nuisances from the acoustic environment. 

Joint factor analysis (JFA) \cite{kenny2007speaker, kenny2008study} was proposed to compensate for the variability issues by separately modeling the inter-speaker variability and the channel or session variability.} The JFA approach employs a GMM supervector, which is a concatenated mean of the adapted GMM. For example, suppose a $F$ by 1 speaker-independent GMM mean vector $m_{c}$, where $c$ is the mixture component index and $F$ is the dimension of the feature. Then, a supervector $\mathbf{M}$ has dimension of $CF$ by 1 by concatenating the F-dimensional mean vector for C mixture components. Thus, the supervector $\mathbf{M}$ can be described as follows:

\begin{align}
\label{eq:mean_supervector}
\mathbf{M} &= \left[m_{1}^{t}, m_{2}^{t}, \ldots, m_{C}^{t}\right]^{t}. 
\end{align}
In the JFA approach, the given GMM supervector is decomposed into speaker independent, speaker dependent, channel dependent, and residual components. {\revision Thus, the ideal speaker supervector $\mathbf{M}_{J}$ can be decomposed as indicated in Eq.~\eqref{eq:JFA}, where $\mathbf{m}_{J}$ denotes a speaker independent supervector from the UBM, $\mathbf{V}$ denotes a speaker dependent component matrix,} $\mathbf{U}$ denotes a channel dependent component matrix, and $\mathbf{D}$ denotes a speaker-dependent residual component matrix. Along with these component matrices, vector $\mathbf{y}$ is for the speaker factors, vector $\mathbf{x}$ is for the channel factors, and vector $\mathbf{z}$ is for the speaker-specific residual factors. All of these vectors have a prior distribution of $N(0,1)$.
\begin{align}
    \label{eq:JFA}
    \mathbf{M}_{J} &= \mathbf{m}_{J} + \mathbf{V y} + \mathbf{U x} + \mathbf{D z}.
\end{align}
{\revision The JFA approach was followed by the study in \cite{Dehak+_2011}, in which it was discovered that channel factors in the JFA also contain information regarding the speakers. Thus, Dehak et al. \cite{Dehak+_2011} proposed a new method combining the channel and speaker spaces into a combined variability space through a \textit{total variability} matrix. Thus, the total variability matrix $\mathbf{T}$ models} both the channel and the speaker variability, and the latent variable $\mathbf{w}$ weights the column of the matrix $\mathbf{T}$. The variable $\mathbf{w}$ is referred to as the \textit{i-vector} and is also considered a speaker representation vector. Each speaker and channel in a GMM supervector $\mathbf{M}_{I}$ can be modeled as follows: 
\begin{align}
    \label{eq:i-vector}
    \mathbf{M}_{I} &= \mathbf{m}_{I} + \mathbf{T w},
\end{align}
{\revision where $\mathbf{m}_{I}$ is a speaker-independent and channel-independent supervector that can be taken as a UBM supervector. The process of extracting an i-vector $\mathbf{w}$  for the given recording is formulated as a MAP estimation problem \cite{kenny2005eigenvoice, Dehak+_2011} using the Baum–Welch statistics extracted using the UBM, mean supervector $\mathbf{m}_{I}$, and total variability matrix $\mathbf{T}$ trained from the EM algorithm as parameters.} The idea behind a speaker representation was greatly popularized through the use of i-vectors, where the speaker representation vector can contain a numerical feature characterizing the vocal tract of each speaker. The i-vector speaker representations have been employed in not only speaker recognition studies but also in numerous speaker diarization studies \cite{senoussaoui2013study, sell2014speaker, zhu2016online} and have shown a superior performance compared to {\revision metric-based methods such as BIC, GLR, and KL, as mentioned in the previous subsection}. 

Intersession variability in the i-vector approach has been further compensated using backend procedures, such as a linear discriminant analysis (LDA) \cite{kanagasundaram2012weighted, kanagasundaram2014vector} and within-class covariance normalization (WCCN) \cite{senoussaoui2010vector, kanagasundaram2011vector}, followed by simple cosine similarity scoring. The cosine similarity scoring was later replaced with a probabilistic LDA (PLDA) model in \cite{matvejka2011full}. In the following studies \cite{garcia2011analysis, kenny2010bayesian}, a method applying a Gaussianization of the i-vectors and thus generating Gaussian assumptions in the PLDA, referred to as G-PLDA or simplified PLDA, was proposed for speaker verification. In general, PLDA employs the following modeling for the given speaker representation $\phi_{ij}$ of the $i$-th speaker {\rrevision in the} $j$-th session as indicated below:

\begin{align}
    \label{eq:PLDA}
    \phi_{i j}=\boldsymbol{\mu}+\mathbf{F h}_{i}+\mathbf{G} \mathbf{w}_{i j}+\epsilon_{i j}.
\end{align}
Here, $\boldsymbol{\mu}$ is the mean vector, $\mathbf{F}$ is the speaker variability matrix, $\mathbf{G}$ is the channel variability matrix, and $\epsilon_{ij}$ is a residual component. In addition, $\mathbf{h}_i$ and $\mathbf{w}_{ij}$ are latent variables {\rrevision specific for the speaker and session, respectively. In G-PLDA, both latent variables, $\mathbf{h}_i$ and $\mathbf{w}_{ij}$, are assumed to follow a standard Gaussian prior. During the training process of the PLDA,  $\boldsymbol{\mu}$, $\mathbf{\Sigma}$, $\mathbf{F}$, and $\mathbf{G}$ are estimated using the expectation maximization (EM) algorithm. Based on the estimated statistics, two hypotheses are tested: hypothesis $H_0$ for a case in which two samples are from the same speaker, and hypothesis $H_1$ for when two samples are from different speakers. Under the hypothesis $H_0$, the given speaker representations $\phi_{1}$ and $\phi_{2}$ are modeled as follows with a common latent variable $\mathbf{h}_{12}$:}
\begin{align}
    \label{eq:hypothesis_H0}
\left[\begin{array}{l}
\phi_{1} \\
\phi_{2}
\end{array}\right]=\left[\begin{array}{l}
\boldsymbol{\mu} \\
\boldsymbol{\mu}
\end{array}\right]+\left[\begin{array}{lll}
\mathbf{F} & \mathbf{G} & 0 \\
\mathbf{F} & 0 & \mathbf{G}
\end{array}\right]\left[\begin{array}{l}
\mathbf{h}_{12} \\
\mathbf{w}_{1} \\
\mathbf{w}_{2}
\end{array}\right]+\left[\begin{array}{l}
\epsilon_{1} \\
\epsilon_{2}
\end{array}\right].
\end{align}
On the other hand, {\revision under the hypothesis $H_1$, $\phi_{1}$ and $\phi_{2}$ are modeled as follows with separate latent variable $\mathbf{h}_1$ and $\mathbf{h}_2$:} 
\begin{align}
    \label{eq:hypothesis_H1}
\left[\begin{array}{c}
\phi_{1} \\
\phi_{2}
\end{array}\right]=\left[\begin{array}{c}
\boldsymbol{\mu} \\
\boldsymbol{\mu}
\end{array}\right]+\left[\begin{array}{cccc}
\mathbf{F} & \mathbf{G} & 0 & 0 \\
0 & 0 & \mathbf{F} & \mathbf{G}
\end{array}\right]\left[\begin{array}{c}
\mathbf{h}_{1} \\
\mathbf{w}_{1} \\
\mathbf{h}_{2} \\
\mathbf{w}_{2}
\end{array}\right]+\left[\begin{array}{c}
\epsilon_{1} \\
\epsilon_{2}
\end{array}\right].
\end{align}
{\revision 
{\rrevision In G-PLDA, it is assumed that $\phi$ is generated from a Gaussian distribution, which results in the following conditional density function \cite{jiang2014plda}.} 
\begin{align}
    \label{eq:Gaussian_prob_plda}
p\left(\phi \mid \mathbf{h}, \mathbf{w}\right)=\mathcal{N}\left(\phi \mid \boldsymbol{\mu}+\mathbf{F h}+\mathbf{G} \mathbf{w}, \mathbf{\Sigma}\right).
\end{align}
{\rrevision Using  Eq.~(\ref{eq:PLDA})-(\ref{eq:hypothesis_H1})}, the log likelihood ratio can be described as follows: }
\begin{align}
    \label{eq:hypothesis_testing}
s\left(\phi_{1}, \phi_{2}\right)=\log p\left(\phi_{1}, \phi_{2} \mid H_{0}\right)-\log p\left(\phi_{1}, \phi_{2} \mid H_{1}\right).
\end{align}
{\revision The log likelihood ratio $s\left(\phi_{1}, \phi_{2}\right)$ in the above equation was originally used for speaker verification to choose a hypothesis between $H_0$ and $H_1$ by checking whether $s\left(\phi_{1}, \phi_{2}\right)$ is positive or negative. \rrevision{The PLDA for speaker representations also employed in speaker diarization} and the log likelihood $s\left(\phi_{1}, \phi_{2}\right)$ is used to check the similarity between clusters. Further details regarding the clustering approach using PLDA is described in Section \ref{subsubsec:AHC}. }

\subsubsection{Neural Network Based Speaker Representations}
\label{sec:Neural Network Based Speaker Representations}

\begin{figure}[t]
  \centering
  \begin{minipage}[b]{\columnwidth}
    \includegraphics[width=\columnwidth]{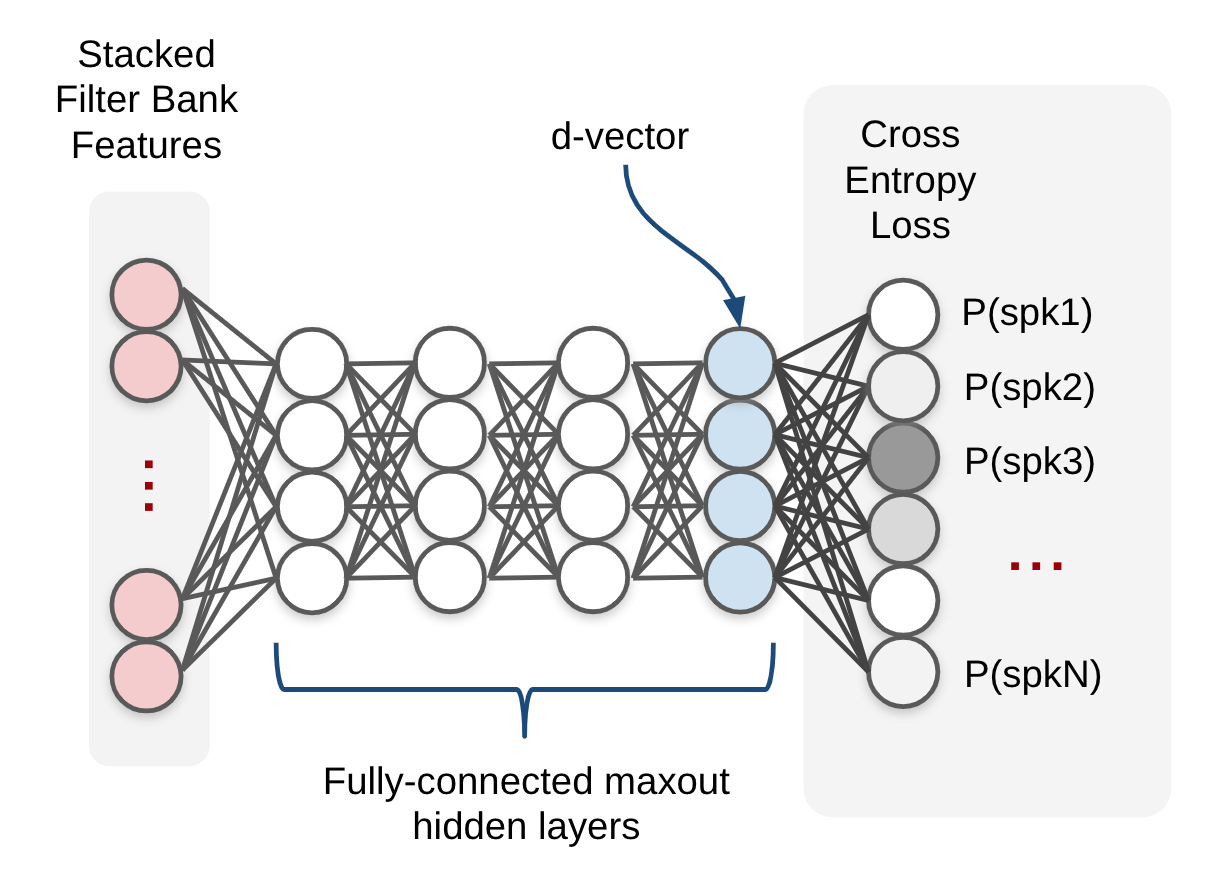}
    \caption{Diagram of d-vector model.}
    \label{fig:d-vector}
  \end{minipage}
  \hfill
  \begin{minipage}[b]{\columnwidth}
    \includegraphics[width=\columnwidth]{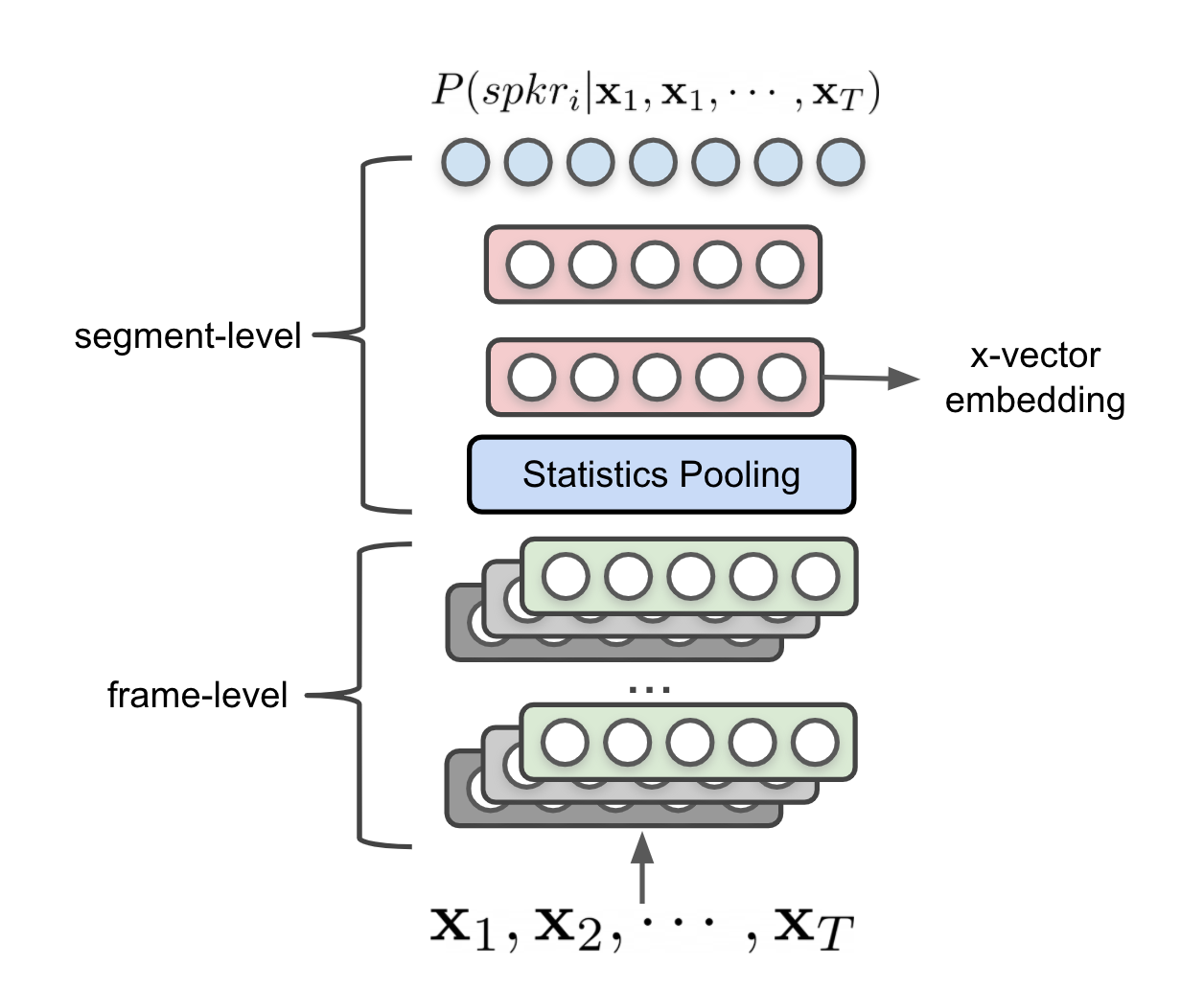}
    \caption{Diagram of x-vector model.}
    \label{fig:x-vector}
  \end{minipage}
\end{figure}

Speaker representations for speaker diarization have also been heavily affected by the rise of deep learning approaches. The idea behind DNN-based representation learning was first introduced for face recognition tasks \cite{sun2014deep, taigman2014deepface}. As the fundamental idea of a neural network-based representation, we can use the deep neural network architecture to map the input signal source (an image or an audio clip) {\revision to a dense vector containing floating-point numbers. This is achieved by taking the values from a layer in the neural network model after forward-propagating the input signal to the layer that we take the values from. 
The mapping process from the input signal to the speaker embedding is based on the nonlinear modeling capability of multiple layers in the DNNs. In so doing, the training process of the DNNs allows the neural networks to learn the mapping without specifying any components or factors, which is in contrast to traditional factor analysis models based on decomposable components. In this sense, the components in JFA are more explainable than the parameters trained in DNN models trained for speaker embedding extraction. In addition, DNN-based speaker representation learning does not involve predefined probabilistic models (e.g., GMM-UBM) for the input acoustic features. In relation to this, DNN-based speaker representation achieves an improved efficiency during the inference phase because the solution used by factor-analysis based methods involves a computationally intensive matrix inversion operation \cite{jiang2014plda}, whereas DNN-based embedding extractors involve fewer demanding operations, such as multiple linear transformations, with non-linear function computations for obtaining the speaker representation vector. Thus, the representation learning process has become more straight-forward and the inference speed has been improved compared to the traditional factor-analysis based methods.} Among many of the neural-network based speaker representations, d-vector \cite{variani2014deep} remains one of the most prominent speaker representation extraction frameworks. The stacked filter bank features, which include context frames as an input feature, are employed, and multiple fully connected layers are trained with cross entropy loss. {\revision Speaker representation vectors, also referred to as d-vectors, are obtained from the last fully connected layer, as indicated in Fig.~\ref{fig:d-vector}.} The d-vector scheme appears in numerous speaker diarization papers, e.g., in \cite{wang2018speaker, zhang2019fully}.
 
DNN-based speaker representations are even more improved when using an x-vector \cite{snyder2017deep, snyder2018x}, which demonstrates a superior performance, winning the NIST speaker recognition challenge 2018 \cite{villalba2019state} and the first DIHARD challenge \cite{sell2018diarization}. Fig.~\ref{fig:x-vector} shows the structure of an x-vector extractor. The time-delay architecture and statistics pooling layer differentiate the x-vector architecture from that of a d-vector. {\revision The statistics pooling layer aggregates the frame-level outputs from the previous layer and computes its mean and standard deviation, passing them on to the following layer. Thus, it can allow the extraction of x-vectors from a variable length input.
This is advantageous not only for speaker verification but also for speaker diarization because speaker diarization systems are bound to process segments that are shorter than the predetermined uniform segment length when the segment should be truncated at the end of an utterance. }

\subsection{Clustering} 
\label{subsec:clustering}
{\tp 
{\revision {\rrevision A} clustering algorithm is applied to make clusters of the speech segments {\rrevision based on the speaker representation and similarity measure explained in the previous section. Here, we} introduce the most commonly used clustering methods for speaker diarization. }

\begin{figure}[t]
  \centering
    \includegraphics[width=\columnwidth]{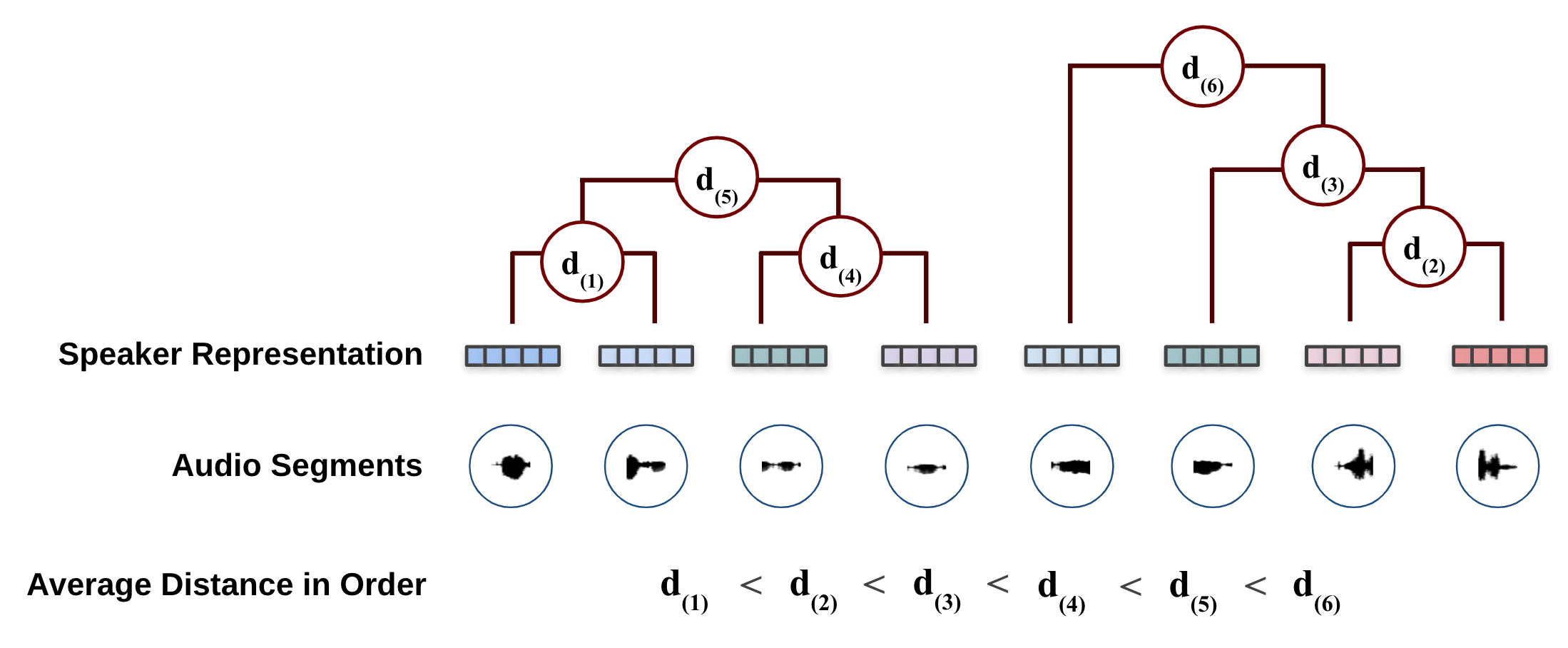}
    \caption{Agglomerative Hierarchical Clustering.}
    \label{fig:ahc}
\end{figure}

\subsubsection{\revision Agglomerative Hierarchical Clustering}
\label{subsubsec:AHC}
AHC is a clustering method that has been constantly employed in many speaker diarization systems with different distance metrics such as BIC \cite{chen1998speaker, han2007robust}, KL \cite{rougui2006fast} and PLDA~\cite{sell2018diarization, arora2020jhu, novoselov2019speaker}. AHC is an iterative process of merging the existing clusters until the clustering process meets a criterion. The AHC process starts with the  calculation of the similarity between N singleton clusters. At each step, a pair of clusters that has the highest similarity is merged. The iterative merging process of AHC is illustrated in a dendrogram, which is presented in Fig.~\ref{fig:ahc}. 

One of the most important aspects of AHC is the stopping criterion. For the speaker diarization task, the AHC process can be stopped using either a similarity threshold or a target number of clusters. Ideally, if PLDA is used as a distance metric, the AHC process should be stopped at $s\left(\phi_{1}, \phi_{2}\right)=0$ in Eq.(\ref{eq:hypothesis_testing}). However, the stopping metric is adjusted to obtain an accurate number of clusters based on a development set. Conversely, if the number of speakers is known or estimated using other methods, the AHC process can be stopped when the clusters created by the AHC process reaches the pre-determined number of speakers $k$.

\subsubsection{Spectral Clustering}
\label{sec:spectral_clustering}

Spectral clustering is a widely used clustering approach for speaker diarization. While there are many variations, spectral clustering involves the following steps.

\begin{enumerate}[label=\roman*.]
    \item Affinity matrix calculation: There are many ways to generate affinity matrix $\mathbf{A}$ depending on the way the affinity value is processed. The raw affinity value $d$ is processed by kernels such as $\exp \left(-d^{2}/\sigma^{2}\right)$, where $\sigma$ is a scaling parameter. On the other hand, the raw affinity value $d$ could also be masked by zeroing the values below a threshold to only keep the prominent values. 
    \item Laplacian matrix calculation \cite{von2007tutorial}: The graph Laplacian can be calculated in two ways: normalized and unnormalized. The degree matrix $\mathbf{D}$ contains diagonal elements  $d_{i}=\sum_{j=1}^{n} a_{i j}$ where $a_{ij}$ is the element of the $i$-th row and $j$-th column in an affinity matrix $\mathbf{A}$.
        \begin{enumerate}
        \item Normalized Graph Laplacian: 
        \begin{align}
            \mathbf{L} &= \mathbf{D^{-1/2}} \mathbf{A} \mathbf{D^{-1/2}}.
        \end{align}
        \item Unnormalized Graph Laplacian: 
        \begin{align}
            \mathbf{L} &= \mathbf{D} - \mathbf{A}.
        \end{align}
        \end{enumerate}
    
    \item Eigen decomposition: The graph Laplacian matrix $\mathbf{L}$ is decomposed into the eigenvector matrix $\mathbf{X}$ and the diagonal matrix that contains eigenvalues. Thus,  $\mathbf{L}=\mathbf{X} \mathbf{\Lambda}\mathbf{X}^{\top}$.
    \item Re-normalization (optional): the rows of $\mathbf{X}$ is normalized so that $y_{i j}=x_{i j} /\left(\sum_{j} x_{i j}^{2}\right)^{1 / 2}$  where $x_{ij}$ and $y_{ij}$ are the elements of the $i$-th row and $j$-th column in matrices $\mathbf{X}$ and $\mathbf{Y}$, respectively. 
    \item Speaker counting: Speaker number is estimated by finding the maximum eigengap \cite{von2007tutorial, park2019auto}.
    \item {\revision Spectral embedding clustering:} The $k$-smallest eigenvalues $\lambda_1$, $\lambda_2$,..., $\lambda_n$ and the corresponding $k$ eigenvectors $\mathbf{v}_1$, $\mathbf{v}_2$,..., $\mathbf{v}_k$ are stacked to construct a matrix  $\boldsymbol{U} \in \mathbb{R}^{n \times k}$. {\revision The row vectors of $\boldsymbol{U}$ are referred to as $k$-dimensional spectral embeddings. Finally, the spectral embeddings are clustered using a clustering algorithm. In general k-means clustering \cite{macqueen1967some} is employed for clustering the spectral embeddings.}
\end{enumerate}

\begin{figure}[t]
  \centering
    \includegraphics[width=\columnwidth]{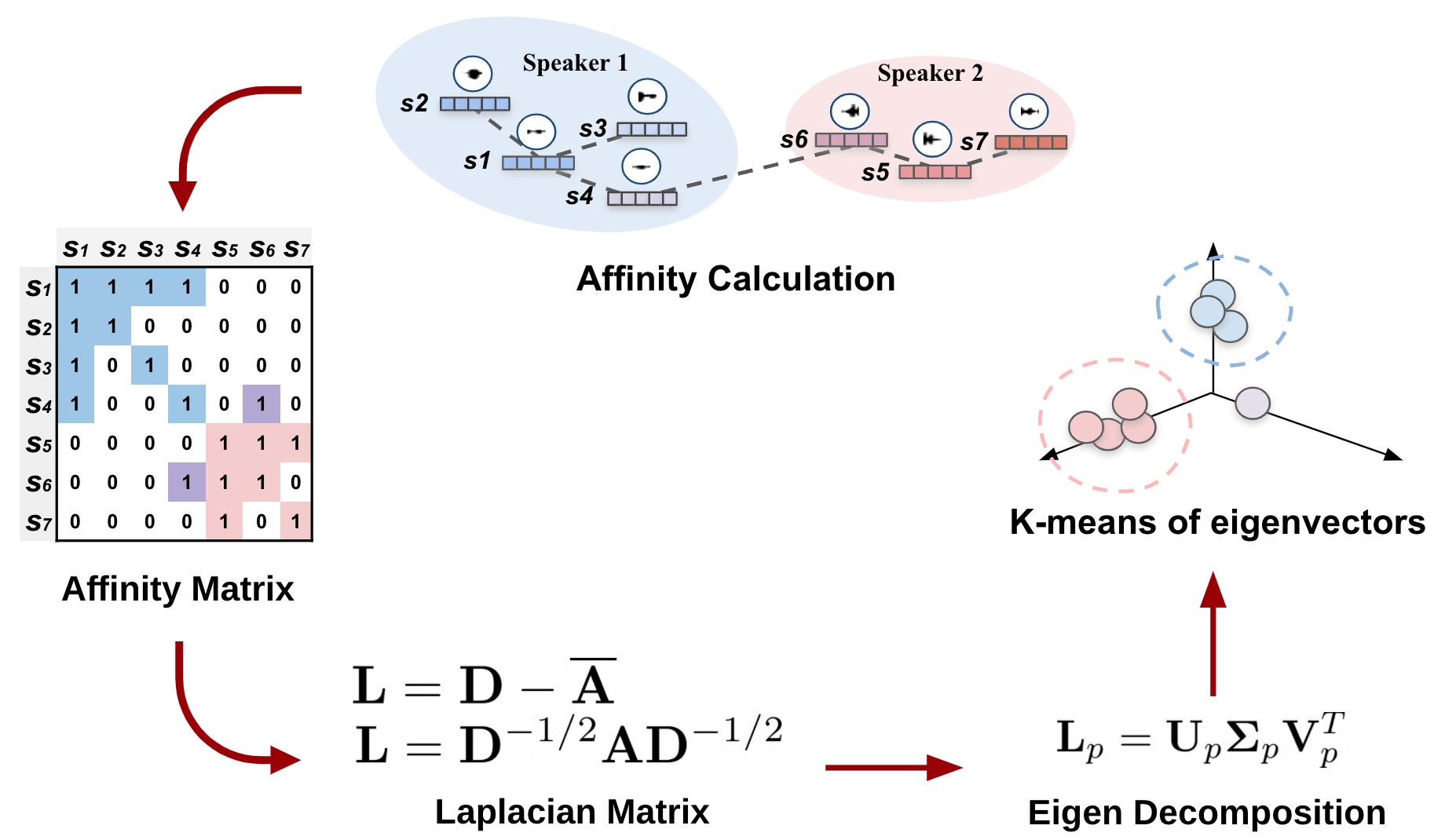}
    \caption{General steps of spectral clustering.}
    \label{fig:sc}
\end{figure}

{\revision 
Among many variations of spectral clustering algorithm, the Ng-Jordan-Weiss (NJW) algorithm \cite{ng2001spectral} is often employed for the speaker diarization task with variation in the kernel for the calculation of the affinity values~\cite{ning2006spectral, luque2012, shum2012use}. Unlike the AHC approach, spectral clustering is mostly used with cosine distance \cite{ning2006spectral, luque2012, shum2012use, wang2018speaker, park2019auto}. In addition, the LSTM based similarity measurement with spectral clustering \cite{lin2019lstm} also exhibited competitive performance. Depending on the datasets, the spectral clustering approach with cosine distance measurement outperforms AHC with PLDA \cite{park2019auto, medennikov2020stc} while using the same speaker representation for both clustering methods. 

\subsubsection{Other Clustering Algorithms}

The k-means algorithm is often employed in studies on speaker diarization \cite{zajic2016investigation, dimitriadis2017developing, wang2018speaker, zhang2019fully, huang2020speaker} due to its simplicity and ease of implementation. However, the k-means algorithm generally underperforms \cite{wang2018speaker, zhang2019fully} the well-known clustering algorithms such as spectral clustering and AHC. In addition, there are a few speaker diarization studies employed the mean-shift \cite{comaniciu2002mean} clustering algorithm, which assigns the given data points to the clusters iteratively by finding the modes in a non-parametric distribution. The Mean-shift clustering algorithm was employed in the speaker diarization task with KL distance in \cite{stafylakis2010speaker}, i-vector and cosine distance in \cite{senoussaoui2013study, senoussaoui2013efficient}, and i-vector and PLDA in \cite{salmun2017plda}. 
}

}



\subsection{Post-processing}
\label{subsec:post-proc}
{\nk 
\subsubsection{Resegmentation}
Resegmentation is a process of refining the speaker boundary
that is roughly estimated using the clustering procedure.
In \cite{kenny2010diarization}, the Viterbi resegmentation method based on the Baum-Welch algorithm
was introduced. In this method, 
the estimation of 
Gaussian mixture model corresponding to
each speaker and
Viterbi-algorithm-based resgmentation using the estimated speaker GMM
are alternately applied.

A method for representing the diarization process based on the variational Bayeian hidden Markov model (VB-HMM) was 
proposed, and was shown to be superior to Viterbi resegmentation \cite{diez2018speaker,diez2019analysis}.
{\revision The VB-HMM-based diarization can be seen as a joint optimization 
of segmentation and clustering, which will be separately introduced 
in
Section \ref{subsec:vb}.}}
\subsubsection{System Fusion}

As another direction of post processing,
there have been a series of studies on 
the fusion method of multiple diarization results to improve the diarization accuracy.
While it is widely known that the system combination generally yields better result for various systems 
(e.g., speech recognition \cite{fiscus1997post} or speaker recognition \cite{brummer2007fusion}),
the combination of multiple diarization hypotheses poses several unique problems.
First, speaker labeling is not standardized among different diarization systems.
Second, the estimated number of speakers may differ among different diarization systems.
Finally, the estimated time boundaries may also be different among multiple diarization systems.
System combination methods for speaker diarization systems need to handle these problems during the fusion process of multiple hypotheses.

In \cite{huijbregts2009majority}, a method for selecting the best diarization result among many  diarization systems {\revision was} proposed.
In this method,
{\revision
a {\rrevision whole sequence of} diarization result {\rrevision for a recording} from each diarization system is
treated as one object to be clustered.
AHC is applied to the set of diarization results, in which the distance of two 
clusters is measured using the symmetric DER between the diarization results belonging to the two clusters.
The iterative merging process of AHC is executed until the number of clusters becomes two.
Finally, in the bigger cluster among the two final clusters {\rrevision according to the number of elements in each cluster}, 
the diarization result that has the smallest
distance to all other diarization results  
is selected as the final result.}
In \cite{bozonnet2010system}, two diarization systems are combined by finding the matching between two speaker clusters, 
and then performing resegementation based on the matching result. 

\begin{figure}[t]
  \centering
  \includegraphics[width=\columnwidth]{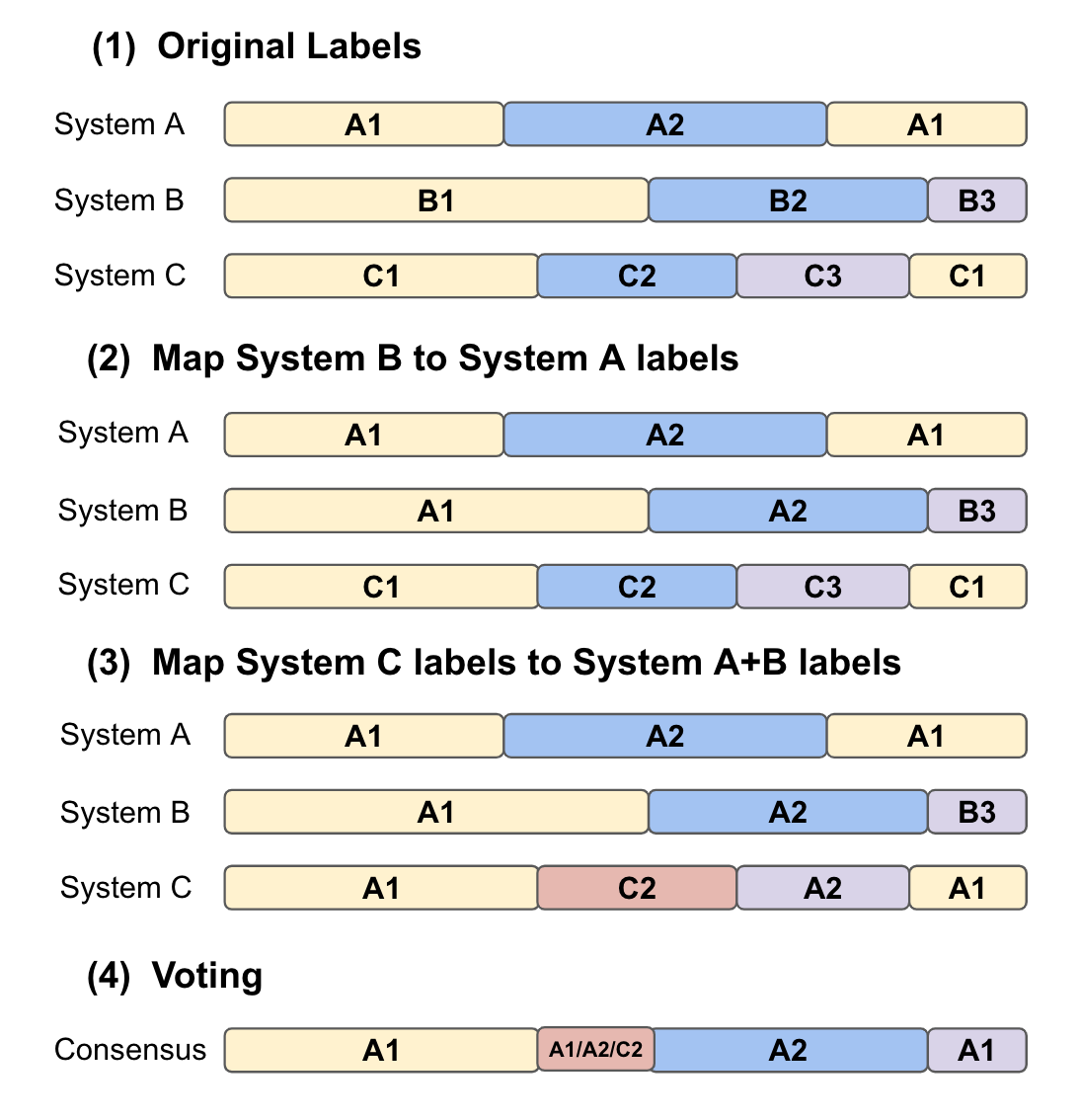}
  \caption{Example of DOVER system.}
  \label{fig:dover}
\end{figure}

More recently,
the diarization output voting error reduction (DOVER) method \cite{stolcke2019dover} was proposed to combine multiple diarization results
based on the voting scheme.
In the DOVER method, the speaker labels among different diarization systems
are aligned one by one to minimize the DER between the hypotheses
(processes 2 and 3 of Fig.~\ref{fig:dover}).
After aligning all hypotheses, each system votes its speaker label to each segmented region (each system may have different weights for
voting), and the speaker label that
gains the highest voting weight is selected for each segmented region (the process 4 of Fig.~\ref{fig:dover}).
In case multiple speaker labels get the same voting weight,
a heuristic approach is employed to break the ties (such as selecting the result from the first system) is used.

The DOVER method has an implicit assumption
that there is no overlapping speech, i.e., at most only one speaker is assigned for each time index.
To combine the diarization hypotheses with overlapping speakers,
two methods were recently proposed. 
In \cite{xiao2020microsoft}, the authors proposed the modified DOVER method, in which the speaker labels in different diarization results
are first aligned with a root hypothesis, 
and the speech activity of each speaker is
estimated based on the weighted voting score for each speaker for each small segment. 
Raj et al. \cite{raj2020dover} proposed a method called DOVER-Lap, in which
the speakers of multiple hypotheses are aligned via a weighted k-partite graph matching, 
and 
the number of speakers $K$ for each small segment is estimated based on the weighted average of
multiple systems to select
the top-$K$ voted speaker labels.
Both the modified DOVER and DOVER-Lap showed DER improvement for the speaker diarization result with speaker overlaps.

\subsection{\revision Joint Optimization of Segmentation and Clustering}
\label{subsec:vb}
{\revision
This subsection introduces a VB-HMM-based diarization technique, which can be regarded
as a joint optimization of segmentation and clustering, and thus cannot be
well categorized in Section 2.1--2.6.
{\rrevision The VB-HMM framework was proposed as an extension of
the VB-based speaker clustering \cite{valente2005variational,kenny2008bayesian} by introducing HMM to constrain the speaker transitions.}
In the VB-HMM framework \cite{diez2018speaker},
the speech feature $\mathbf{X}=(\mathbf{x}_t|t=1,...,T)$ is assumed to be generated from HMM where
each HMM state corresponds to one of $K$ possible speakers.
Suppose that we have $M$ HMM states, $M$-dimensional variable $\mathbf{Z}=(\mathbf{z}_t|t=1,...,T)$ is introduced where
$k$-th element of $\mathbf{z}_t$ is 1 if the $k$-th speaker is speaking at the time index $t$, and 0 otherwise.
At the same time, the distribution of $\mathbf{x}_t$ 
is modeled based on a hidden variable $\mathbf{Y}=\{\mathbf{y}_k|k=1,...,K\}$,
 where $\mathbf{y}_k$ denotes a low dimensional vector
for the $k$-th speaker. 
Given these notations,
the joint probability of $\mathbf{X}$, $\mathbf{Y}$, and $\mathbf{Z}$ is decomposed as follows:
\begin{equation}
P(\mathbf{X},\mathbf{Z},\mathbf{Y})=P(\mathbf{X}|\mathbf{Z},\mathbf{Y})P(\mathbf{Z})P(\mathbf{Y}),
\end{equation}
where $P(\mathbf{X}|\mathbf{Z},\mathbf{Y})$ is the emission probability modeled by GMM whose mean vector is represented by $\mathbf{Y}$, $P(\mathbf{Z})$ is the transition probability of the HMM, and
$P(\mathbf{Y})$ is the prior distribution of $\mathbf{Y}$.
Because $\mathbf{Z}$ represents the trajectory of speakers,
the diarization problem can be expressed as the inference problem of $\mathbf{Z}$ that maximizes 
the posterior distribution $P(\mathbf{Z}|\mathbf{X})=\int P(\mathbf{Z},\mathbf{Y}|\mathbf{X})d\mathbf{Y}$.
Since it is intractable to directly solve this problem, the VB method is used to estimate
the model parameters that approximate $P(\mathbf{Z},\mathbf{Y}|\mathbf{X})$.

Recently, a simplified version of VB-HMM that works on the x-vector, known as VBx, 
was proposed \cite{diez2019bayesian,landini2020vbx}.
In VBx, $P(X|Z,Y)$ is calculated using the x-vector based on the PLDA model.
While the original VB-HMM works on the granularity of the frame-level feature, 
VBx works on the granularity of the x-vector, and thus can be seen as a clustering method that 
jointly models speaker turn and speaker duration.

The VB-HMM diarization was originally designed as a standalone diarization framework.
However, it requires parameter initialization to start the VB estimation, and the parameters are usually initialized
based on the result of
another speaker clustering.
In that context, VB-HMM is widely employed as the final step of speaker diarization (e.g., \cite{sell2015diarization,Sell18,diez2020optimizing}).
For example in \cite{diez2020optimizing}, AHC was first performed to under-cluster the x-vector,
and VBx was then applied to obtain a better cluster given the AHC-based result as the initial parameter.
Finally, VB-HMM was further applied to refine the boundary obtained by VBx.}

{
\setlength{\tabcolsep}{1mm}
\begin{table*}[t]
\centering
\caption{\revision Overview of speaker diarization techniques using deep learning. Note that there are also a lot of studies that use deep learning for front-end, SAD, segmentation, and speaker embedding extraction, which are introduced in Section \ref{Conventional Speaker Diarization Systems}. }
\vspace{0.5ex}
\label{tab:diarization_techniques}
{
\small
\begin{tabular}{c|cccccc|c|cccc|l} \hline
Section&Front-& SAD   & Seg.  & Speaker&Clust.& Post- & Joint   & Speaker  &Overlap &Inference&Max \# of & \multicolumn{1}{c}{References} \\
      & end   &       &       & embed. &      & proc. & w/ ASR  & profile  &handling& mode    & speakers&                              \\ \hline

3.1.1 & -     & -     & -     & -     &$\surd$& -     & -       & -        & no    & offline &Variable   &Affinity matrix \cite{wang2020speaker,lin2020self,park2020multi}, IDEC \cite{Dimitriadis19} \\ 
3.1.2 & -     & -     & -     & -     &$\surd$& -     & -       & required & no    & online  &Fixed      &RRNN \cite{FlDi20} \\  
3.1.3 & -$^\star$     & -     & -     & -     & -     &$\surd$& -       & required & yes   & offline &Fixed      &TS-VAD$^{\dagger}$ \cite{medennikov2020stc,medennikov2020target} \\ 
3.1.3 & -$^\star$     & -     & -     & -     & -     &$\surd$& -       & -        & yes   & offline &Variable   &EEND-based post-processing$^{\ddagger}$ \cite{horiguchi2020post} \\ \hline
3.2.1 & -     & -     &$\surd$& -     &$\surd$& -     & -       & -        & no    & both    &Variable   &UIS-RNN \cite{zhang2019fully} \\
3.2.2 & -$^\star$     & -     &$\surd$&$\surd$&       &$\surd$& -       & -        & yes   & offline &Variable   &RPN  \cite{huang2020speaker} \\
3.2.3 &$\surd$& -     &$\surd$&$\surd$&$\surd$& -     & -       & -        & yes   & online  &Variable   &Online-RSAN \cite{von2019all,kinoshita2020tackling} \\
3.2.4 & -$^\star$     &$\surd$&$\surd$&$\surd$&$\surd$& -     & -       & -        & yes   & offline &Fixed      &EEND \cite{fujita2019end,fujita2019end2}\\ 
3.2.4 & -$^\star$     &$\surd$&$\surd$&$\surd$&$\surd$& -     & -       & -        & yes   & offline &Variable   &EEND-EDA \cite{horiguchi2020end}, SC-EEND \cite{fujita2020neural} \\ 
3.2.4 & -$^\star$     &$\surd$&$\surd$&$\surd$&$\surd$& -     & -       & -        & yes   & online  &Fixed      & EEND with speaker tracing buffer \cite{xue2020online}\\
3.2.4 & -$^\star$     &$\surd$&$\surd$&$\surd$&$\surd$& -     & -       & -        & yes   & online  &Variable   &BW-EDA-EEND \cite{han2020bw} \\
3.2.4 & -$^\star$     &$\surd$&$\surd$&$\surd$& -     & -     & -       & -        & yes   & offline &Variable   &EEND-vector clustering \cite{kinoshita2020integrating} \\ \hline
4.3   & -     & -     &$\surd$& -     &$\surd$& -     & $\surd$ & -        & no    & offline &Fixed      &Speaker-tag insertion \cite{Shafey2019,mao2020speech} \\
4.3   & -$^\star$     & -     &$\surd$& -     &$\surd$& -     & $\surd$ & -        & yes   & offline &Variable   &MAP-decoding  \cite{kanda2019simultaneous} \\
4.3   & -$^\star$     & -     &$\surd$&$\surd$&$\surd$& -     &$\surd$  & required & yes   & offline &Variable   &End-to-end SA-ASR  \cite{kanda2020joint,kanda2020minimum} \\
4.3   & -$^\star$     & -     &$\surd$&$\surd$& -     & -     & $\surd$ & -        & yes   & offline &Variable   &End-to-end SA-ASR+Clustering  \cite{kanda2021investigation} \\
\hline
\end{tabular}
{
\begin{flushleft}
$^\dagger$ TS-VAD can also be interpreted as a joint model of SAD, segmentation, and speaker identification.
\\$^\ddagger$ EEND \cite{fujita2019end,fujita2019end2} itself is a joint model of SAD, segmentation, speaker embedding, and clustering.
\\$^\star$ It can also be seen that a speech separation module is implicitly embedded in the model to cope with speaker 
overlaps.

\end{flushleft}
}
}
\end{table*}
}

\section{Recent Advances in Speaker Diarization Using Deep Learning}
\label{Recent Neural Approaches in Speaker Diarization}

{\nk
This section introduces various recent efforts 
toward deep learning-based speaker diarization techniques.
The methods that incorporate deep learning into a single component of speaker diarization, such
as clustering or post-processing, are introduced
in Section \ref{sec:non-joint-diarization-optimization}. 
The methods that unify several components of speaker diarization into a single neural network
are introduced in Section \ref{sec:joint-diarization-optimization}{\revision. For the overview of speaker diarization techniques using deep learning, refer to Table \ref{tab:diarization_techniques}. It should be noted that there are some works that take additional input of speaker profiles. These methods may not be categorized as a diarization technique in a traditional definition. Nevertheless, we introduce them as they are 
 optimized in a multispeaker situation to learn the relations between speakers and hence categorized as ``Trained Based on the Diarization Objective'' in Table \ref{tab:TableOfTaxonomy}.}
}

\subsection{Single-module Optimization}

\label{sec:non-joint-diarization-optimization}

\subsubsection{Speaker clustering Enhanced by Deep Learning}
\label{subsec:clustering_by_dl}

\begin{figure}[t]
  \centering
    \includegraphics[width=\columnwidth]{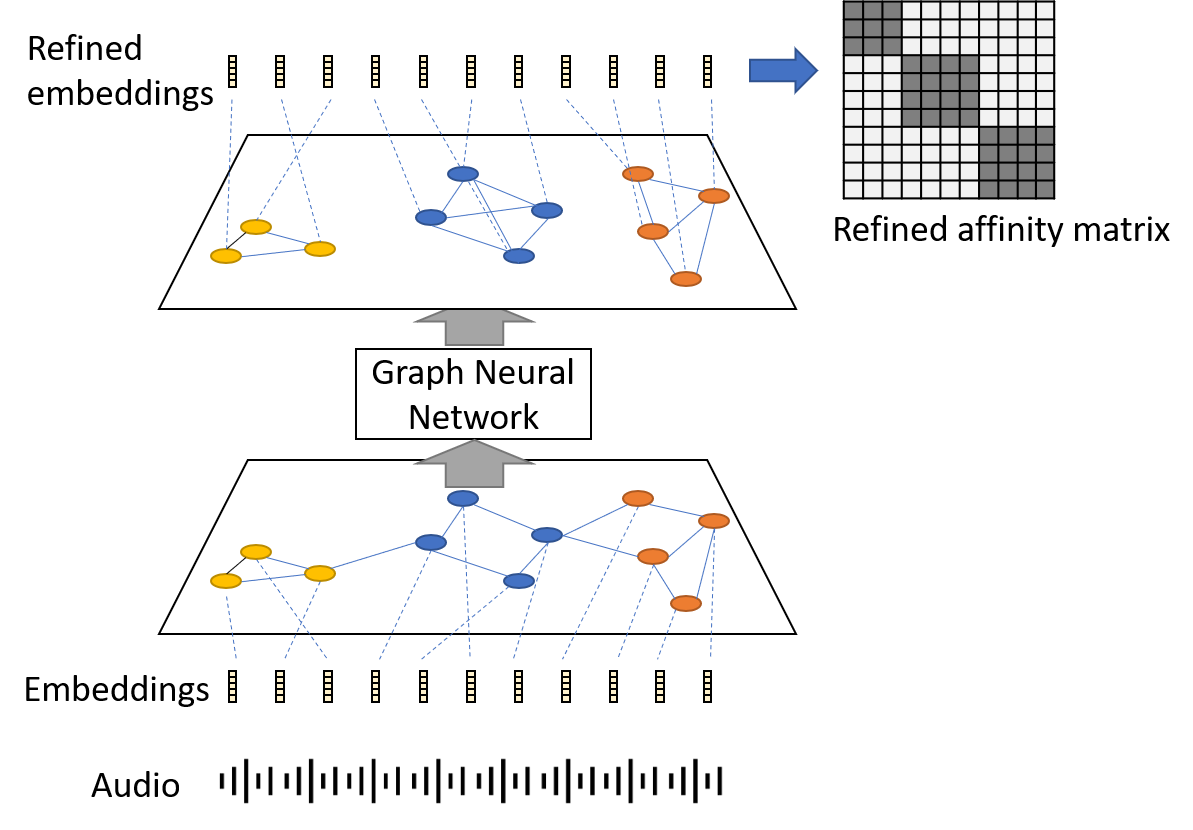}
  \caption{Speaker diarization with graph neural network}
  \label{fig:gnn-diarization}
\end{figure}

{\rrevision Enhancing the clustering procedure based on the deep learning is an active research area and several methods have been proposed for speaker diarization. This section will cover the representative works in such a direction. 

An approach based on the graph neural network (GNN) was proposed in \cite{wang2020speaker}. As shown in Fig.~\ref{fig:gnn-diarization}, this method aims at purifying the similarity matrix used in the spectral clustering (Section \ref{sec:spectral_clustering}).  Assuming a sequence of speaker embeddings $\{\mathbf{e}_1,...\mathbf{e}_N\}$, where $N$ is the sequence length, the input $\mathbf{x}_i^0$ to  the GNN is $\{\mathbf{x}_{i}^0=\mathbf{e}_i|i=1,\ldots,N\}$. The  output $\mathbf{x}_i^{(p)}$ of the $p$-th layer of the GNN is now:
\begin{align}
   x_i^{(p)}= \sigma (\mathbf{W}\sum_j \mathbf{L}_{i,j}x_j^{(p-1)}),
\end{align}
where $\mathbf{L}$ is a normalized affinity matrix added by self-connection, $\mathbf{W}$ is a trainable weight matrix for the $p$-th layer, and $\sigma(\cdot)$ is a nonlinear function. The GNN is trained by minimizing the distance between the reference and the estimated affinity matrices. The distance is calculated using a combination of histogram loss \cite{ustinova2016learning} and nuclear norm \cite{recht2010guaranteed}.  The GNN-based speaker diarization method was evaluated on the CALLHOME dataset and an in-house meeting dataset, and  significantly outperformed any of the conventional clustering methods. 

Besides that, different approaches have been proposed to generate the affinity matrix. In \cite{lin2020self}, a self-attention-based neural network model was introduced to directly generate a similarity matrix from a sequence of speaker embeddings. In \cite{park2020multi}, several affinity matrices with different temporal resolutions were fused into a single affinity matrix based on a neural network.

A different approach aiming at  improving clustering was proposed in~\cite{XGF2016},  called deep embedded clustering (DEC).  The goal of DEC was to transform the input features (herein referred to as speaker embeddings) making them more separable in a given number of clusters/speakers. In order to make cluster differentiable,  each embedding is provided with a probability of ``belonging'' to each of the available speaker clusters, {\revision i.e. $q_{i,j}$ can be interpreted as the probability of assigning sample $i$ to cluster $j$ (i.e., a soft assignment):} 
\begin{equation}
    q_{ij}=\frac{ \left(1+\| z_i-\mu_j\|^2/a\right)^{-\frac{a+1}{a}} }{\sum_l\left(1+\| z_i-\mu_l\|^2/a\right)^{-\frac{a+1}{a}}},\ \ p_{ij}=\frac{q^2_{ij}/f_i}{\sum_l q^2_{il}/f_l},
    \label{Eq:DEC}
\end{equation}
where $z_i$ is the bottleneck feature, {\revision $a $ is the degree of freedom of the Student's $t-$distribution, } $\mu_i$ is the centroid of $i$-th cluster and $f_i$ is the soft cluster frequency with $f_i=\sum q_{ij}$. 
The clusters are iteratively refined based on the target distribution according to the bottleneck features estimated using an autoencoder.

The initial version of DEC had some problems, and refined algorithm called improved DEC (IDEC) was later proposed with better accuracy on speaker diarization \cite{GGLY2017,Dimitriadis19}. Firstly, there was a potential risk that the neural network is converged to a trivial solution to generates corrupted embeddings. To avoid this risk, Guo et al.~\cite{GGLY2017} proposed to explicitly preserve the local structure of the data by adding a reconstruction loss between the output of the autoencoder and the input feature. Dimitriadis~\cite{Dimitriadis19} further addressed the issue by introducing the loss function to enforce the distribution of speaker turns being uniform across all speakers, i.e., all speakers contribute equally to the session. This assumption is not always valid for real recordings but it constrains the solution space enough to avoid the empty clusters without affecting the overall performance. Finally, Dimitriadis~\cite{Dimitriadis19} also proposed an additional loss term that  penalizes the distance from the centroid $\mu_i$, bringing the behavior of the algorithm closer to k-means. 

Overall, 
the loss function of the IDEC consists of four loss terms, i.e., $L_c$, the clustering error term that is originally proposed in DEC; {\revision $L_r$, the reconstruction error term~\cite{GGLY2017}; $L_u$ is the uniform ``speaker airtime'' distribution loss~\cite{Dimitriadis19}; and $L_{MSE}$, the loss to measure the distance of the bottleneck features from the centroids~\cite{Dimitriadis19}},
\begin{equation}
    L= \alpha L_c + \beta L_r + \gamma L_u + \delta L_{MSE},
    \label{eq:loss}
\end{equation}
where $\alpha, \beta, \gamma, \text{ and } \delta$ are the weight on the loss functions that is fine-tuned on some held-out data.

}

\subsubsection{Learning the Distance Estimator}

{\dd
{\rrevision In this section a novel approach using a trainable distance function is presented. The basic idea is based on the relational recurrent neural networks (RRNNs).}
RRNNs were introduced by~\cite{santoro2018relational, santoro2016, sukhbaatar2015end} to address ``relational information learning" problems. {\rrevision Such models learn relations between a sequence of input features like the notion of ``closer'' or ``further'', e.g, two points in space are closer than a third one, etc.}
Speaker diarization can be seen as part of  this class of problems, since the final decision depends on the distance between speech segments and speaker profiles or centroids. 

{\rrevision There are several issues that potentially limits the accuracy of speaker diarization systems. Firstly, as mentioned in Section~\ref{subsec:segmentation}, the duration of segments when extracting speaker embeddings poses a trade-off between the time resolution and the robustness of the extracted speaker representations.
Secondly,  speaker embedding extractors are  not explicitly trained to provide optimal representations for speaker diarization, despite the fact these invariant,  discriminative representations are used to separate thousands of speakers~\cite{snyder2018x}. 
Thirdly, the distance metric is often based on a heuristic approach and/or dependent on certain assumptions that do not necessarily hold, e.g., assuming Gaussianity in the case of PLDA~\cite{garcia2011analysis}. Finally, the audio chunks are treated independently and any temporal context is simply ignored in conventional clustering methods as described in Section~\ref{subsec:clustering}. These issues can be attributed to the distance metric function, and most of them can be addressed with RRNNs, where a data-driven, memory-based approach bridges the performance gap between the heuristic and the trainable distance estimation approaches.

In this context, an} approach for learning the relationship between the speaker cluster centroids (or speaker profiles) and the embeddings is proposed in~\cite{FlDi20} (Fig. \ref{fig:baseline_nthfar}). In this work, the diarization process is considered to be a classification task on an already segmented audio, as in Section~\ref{subsec:segmentation},  either uniformly~\cite{zajic2016investigation} or based on estimated speaker-change salient points~\cite{yoshioka2019meeting}. The speaker embeddings $x_j$ for each segment, which are assumed to be speaker-homogeneous, are extracted and then compared with all the available speaker profiles or speaker centroids. The most suitable speaker label is assigned to each segment by minimizing the distance-based loss function, i.e., the  relationship between embeddings and profiles. 
As discussed in~\cite{FlDi20}, the RRNN-based distance estimation exhibits consistent improvements in its performance when compared with the more traditional distance estimation approaches such as the cosine distance~\cite{Dehak+_2011} or the PLDA-based~\cite{garcia2011analysis} distance.
{\rrevision Note that
although the task in~\cite{FlDi20} is speaker identification, an extension to the speaker diarization is rather straightforward when the speaker profiles are pre-estimated, either as centroids using any \dd{of the} traditional clustering algorithm (Section~\ref{subsec:clustering} and~\ref{subsec:clustering_by_dl}), or using prior knowledge. }

\begin{figure}[t]
  \centering
  \includegraphics[width=\columnwidth]{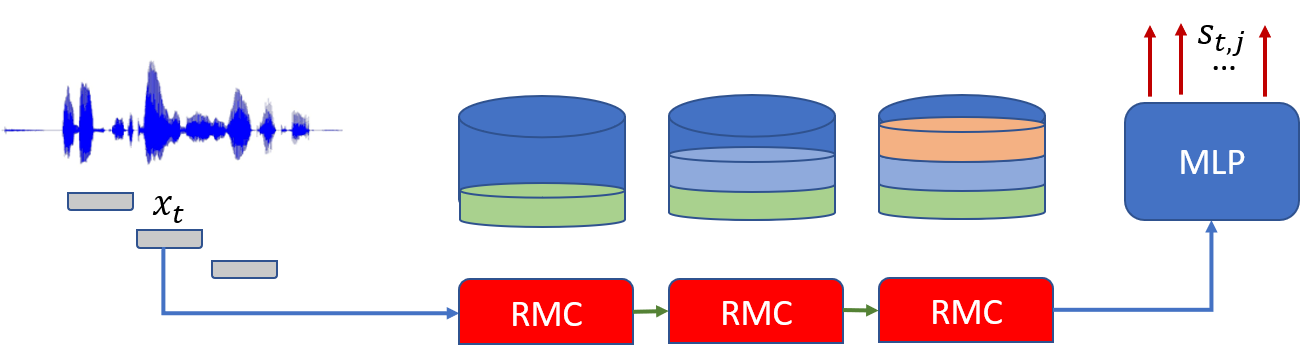}
  \caption{Continuous speaker identification system based on RMC. The speech signal is segmented uniformly and each segment $x_t$ is compared against all the available speaker profiles  according to a distance metric $d(\cdot,\cdot)$. A speaker label $s_{t,j}$ is assigned to each $x_t$ minimizing this metric.}
  \label{fig:baseline_nthfar}
\end{figure}

}

\subsubsection{Post Processing Based on Deep Learning}

\begin{figure}[t]
  \centering
    \includegraphics[width=\columnwidth]{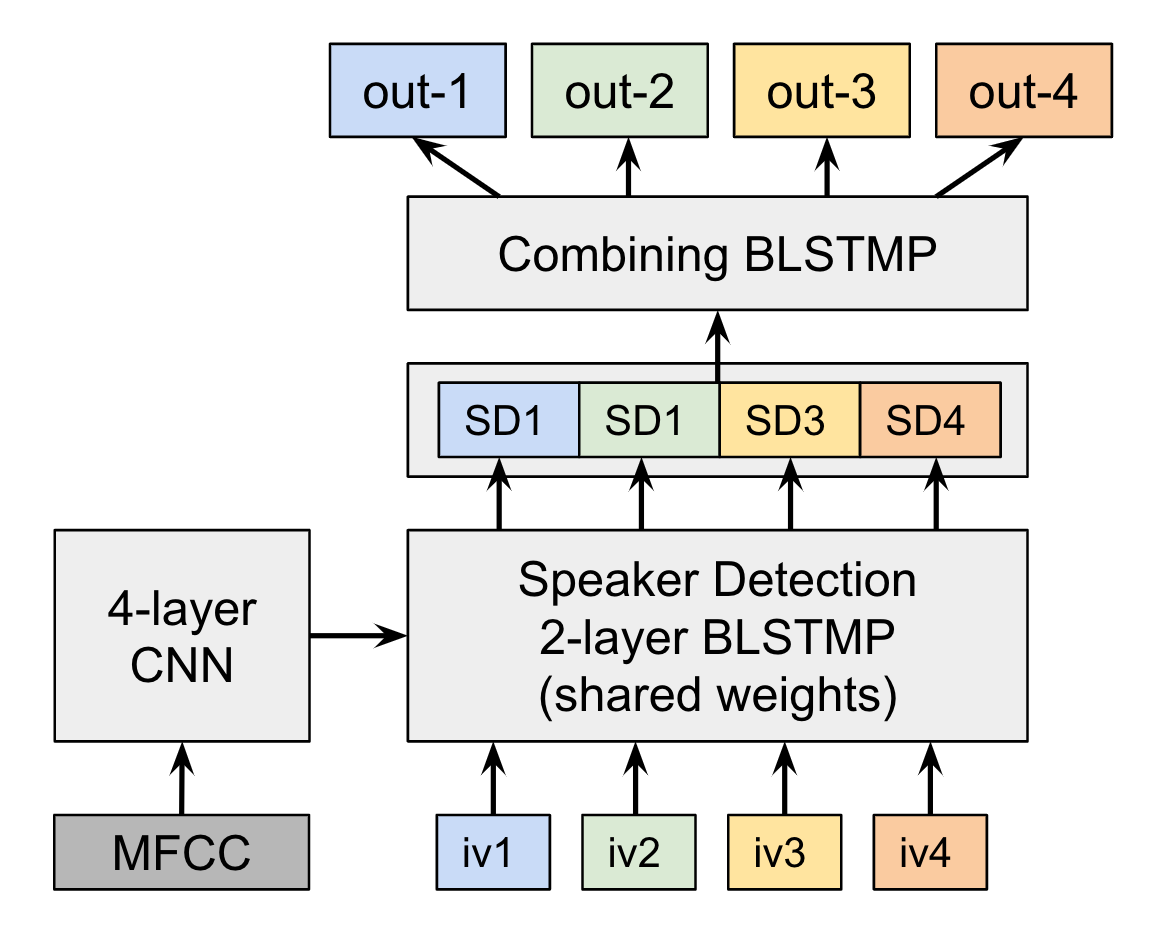}
  \caption{Target Speaker Voice Activity Detector}
  \label{fig:tsvad}
\end{figure}

{\nk
There are a few recent studies on the neural network-based diarization method that is applied on top of the result from a traditional clustering-based speaker diarization. These methods can be categorized as an extension of the post-processing. Medennikov et al. \cite{medennikov2020stc,medennikov2020target} proposed the target-speaker voice activity detection (TS-VAD) to achieve accurate speaker diarization even {\revision under noisy conditions with many speaker overlaps.
TS-VAD assumes that 
a set of i-vectors 
$\mathcal{E}=\{\mathbf{e}_k\in\mathbb{R}^{f}|k=1,...,K\}$
are available for each speaker in the audio,
where 
 $f$ is the dimension of i-vector
and $K$ is the number of speakers.
As presented in Fig. \ref{fig:tsvad}, 
TS-VAD takes 
not only 
 a sequence of MFCC, 
 $\mathbf{X}=(\mathbf{x}_t\in\mathbb{R}^{d}|t=1,...,T)$, where $d$ is the dimension of MFCC and $T$ the length of the sequence, 
 but also 
a set of i-vectors $\mathcal{E}$. 
Given $\mathbf{X}$ and $\mathcal{E}$, the model outputs a sequence of $k$-dimensional vector
$\mathbf{O}=(\mathbf{o}_t\in\mathbb{R}^K|t=1,...,T)$
where the $k$-th element of $\mathbf{o}_t$ 
represents the probability of the speech activity 
of the 
speaker corresponding to $\mathbf{e}_k$ at the time frame $t$.
In other words,
the $k$-th element of $\mathbf{o}_t$ is expected to be 1 if the speaker 
of $\mathbf{e}_k$
 is speaking at time $t$, and 0 otherwise.

Because TS-VAD requires the i-vectors of speakers,
pre-processing to obtain the i-vectors is necessary.
The procedure proposed in \cite{medennikov2020stc,medennikov2020target}
is as follows:
\begin{enumerate}
    \item Apply clustering-based diarization.
    \item Estimate i-vectors for each speaker given the diarization result.
    \item Repeat (a) and (b).
\begin{enumerate}
    \item Apply TS-VAD given the estimated i-vectors.
    \item Refine i-vectors given the TS-VAD result.
\end{enumerate}
\end{enumerate}
TS-VAD was proposed as a part of the winning system of CHiME-6 Challenge \cite{watanabe2020chime},
and showed a significantly better DER compared with the conventional clustering
based approach \cite{medennikov2020stc}.
However, it has a drawback, i.e., the maximum number of speakers
that the model can handle is limited by the dimension of the output vector.}

As a different approach,
Horiguchi et al. proposed the application of the EEND model
(detailed in Section \ref{End-to-End Neural Diarization})
to refine the result of a clustering-based speaker diarization 
\cite{horiguchi2020post}.
A clustering-based speaker diarization method can handle a large number of speakers but 
unable to handle overlapped speech.
Conversly, EEND has opposite characteristics.
To complementarily use the two methods,
The authors in \cite{medennikov2020stc} first applied a conventional clustering method.
Then, the two-speaker EEND model was iteratively applied for each pair of detected speakers 
to refine the time boundary of overlapped regions.
}

\subsection{Joint Optimization for Speaker Diarization}

\label{sec:joint-diarization-optimization}
\subsubsection{Joint Segmentation and Clustering}

{\nk
A model called unbounded interleaved-state recurrent neural networks (UIS-RNN) was proposed, which replaced the segmentation and clustering methods with a trainable model
\cite{zhang2019fully}.
Given the input sequence of embeddings $\mathbf{X}=(\mathbf{x}_t\in \mathbb{R}^d|t=1,\ldots,T)$,
UIS-RNN generates the diarization result $\mathbf{Y}=(y_t\in \mathbb{N}|t=1,\ldots,T)$ as a sequence of speaker index for each time frame.
The joint probability of $\mathbf{X}$ and $\mathbf{Y}$ can be decomposed by the chain rule as follows.
\begin{align}
    P(\mathbf{X},\mathbf{Y})=& P(\mathbf{x}_1,y_1)\prod_{t=2}^T P(\mathbf{x}_t,y_t|\mathbf{x}_{1:t-1},y_{1:t-1}).
\end{align}
To model the distribution of the speaker change,
UIS-RNN then introduces a latent variable
    $\mathbf{Z}=(z_t\in\{0,1\}|t=2,\ldots,T)$, where $z_t$ becomes 1 if
    the speaker indices at time $t-1$ and $t$ are different, and 0 otherwise.
The joint probability including $\mathbf{Z}$ is then decomposed as follows.
\begin{align}
    P(\mathbf{X},\mathbf{Y},\mathbf{Z})=& P(\mathbf{x}_1,y_1)\prod_{t=2}^T P(\mathbf{x}_t,y_t,z_t|\mathbf{x}_{1:t-1},y_{1:t-1},z_{1:t-1}). 
\end{align}
Finally, the term $P(\mathbf{x}_t,y_t,z_t|\mathbf{x}_{1:t-1},y_{1:t-1},z_{1:t-1})$ is further decomposed into three components.
\begin{align}
P(\mathbf{x}_t,y_t,z_t|\mathbf{x}_{1:t-1},y_{1:t-1},z_{1:t-1})  =& P(\mathbf{x}_t|\mathbf{x}_{1:t-1},y_{1:t})P(y_t|z_t,y_{1:t-1})P(z_t|z_{1:t-1}).
\end{align}
Here, 
$P(\mathbf{x}_t|\mathbf{x}_{1:t-1},y_{1:t})$ denotes the sequence generation probability,
and modeled by gated recurrent unit (GRU)-based recurrent neural networks.
$P(y_t|z_t,y_{1:t-1})$ denotes the speaker assignment probability 
and modeled by a distance-dependent Chinese restaurant process \cite{blei2011distance},
which can model the distribution of unbounded number of speakers.
Finally,
$P(z_t|z_{1:t-1})$ represents the speaker change probability and
is modeled by the Bernoulli distribution.
Since all models are represented by trainable ones,
UIS-RNN can be trained in a supervised fashion
by finding the parameters that maximize $\log  P(\mathbf{X},\mathbf{Y},\mathbf{Z})$
over training data.
The inference can be conducted by finding
$\mathbf{Y}$ that maximizes $\log P(\mathbf{X},\mathbf{Y})$ given $\mathbf{X}$ based on the 
beam search in an online fashion.
While UIS-RNN works in an online fashion, it demonstrated better DER than that of the offline system based on spectral clustering.

}

\subsubsection{Joint Segmentation, Embedding Extraction, and Re-segmentation}

\begin{figure}[t]
  \centering
  \includegraphics[width=\columnwidth]{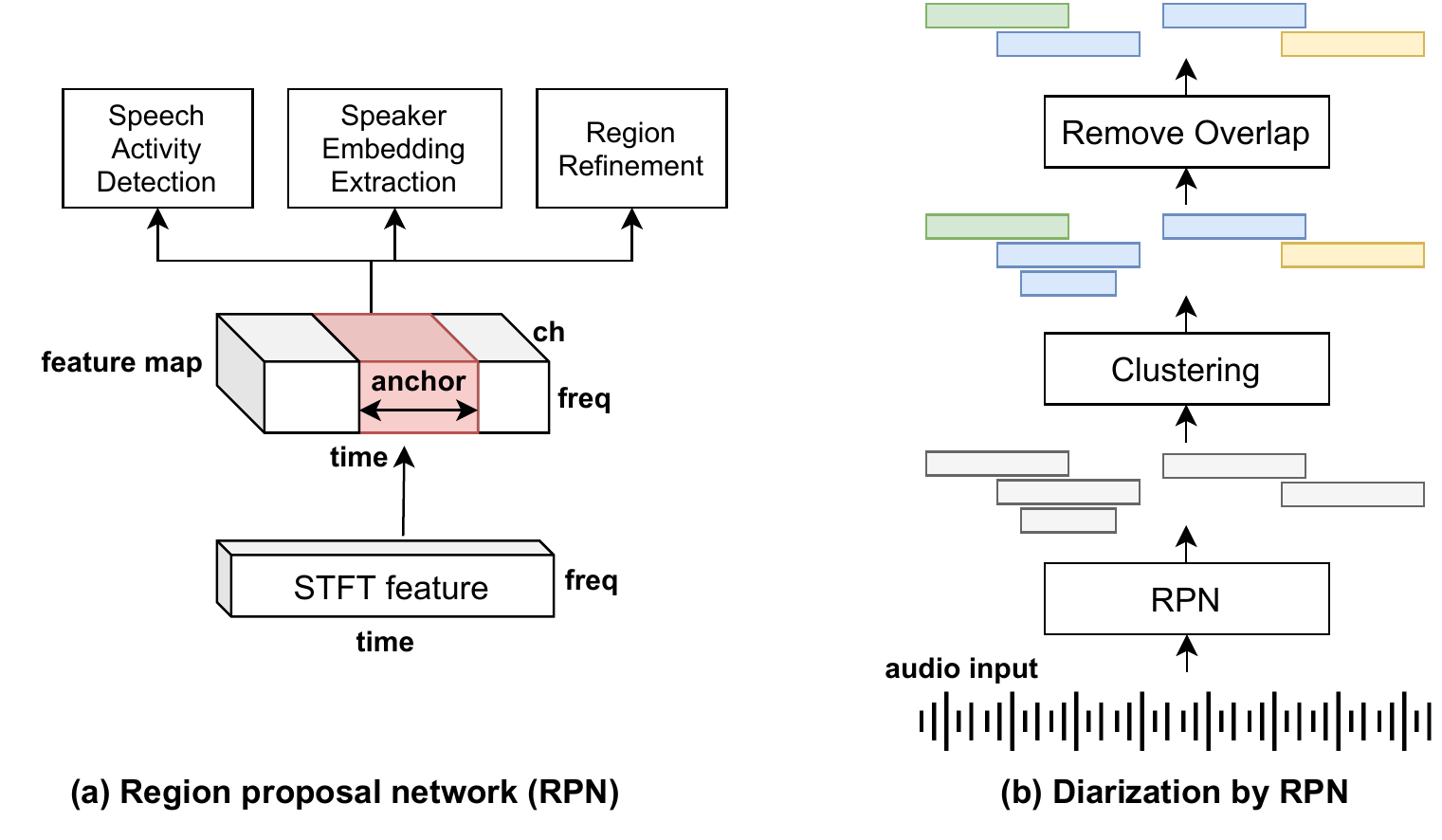}
  \caption{(a) RPN for speaker diarization, (b) diarization procedure based on RPN.}
  \label{fig:rpn}
\end{figure}

{\nk
A speaker diarization method based on the region proposal networks (RPN) was proposed
to jointly perform
segmentation, speaker embedding extraction, and resegmentation \cite{huang2020speaker}.
The RPN was originally proposed to detect multiple objects from a two-dimensional image \cite{ren2016faster},
and one-dimensional variant {\revision along with the time-axis} is used for speaker diarization.

{\revision As can be seen from Fig. \ref{fig:rpn} (a), the STFT features with 
a size of time and frequency bin
is first converted to the 
feature map with a size of time, frequency and channels
using CNNs.
Then, other three types of neural networks
are applied on various sizes of sliding windows (named ``anchor'') along with the time axis.
For each anchor, the three neural networks
perform SAD, speaker embedding extraction, and region
refinement, respectively.
Here, SAD is the task to estimate the probability of
speech activity for the anchor region.
Speaker embedding extraction is the task to generate an embedding to represent
the speaker characteristics of the audio corresponding to the anchor region.
Finally, region refinement is the task to estimate the difference between
the shape (i.e. duration and center position) of the anchor and that of 
the corresponding reference region.}

The inference procedure by RPN is
{\revision presented in Fig. \ref{fig:rpn} (b).
RPN is first applied to list the anchors} with speech activity probability higher than the pre-determined threshold.
{\revision The anchors} are then clustered using a conventional clustering method (e.g., k-means) 
based on the speaker embeddings {\revision estimated for each anchor}.
Finally, {\revision highly overlapped anchors after region refinement are removed, a method known as the non-maximum suppression}.

The RPN-based speaker diarization system has the advantage of handling overlapped speech with possibly any number of speakers.
Also, it is much simpler than the conventional speaker diarization system.
It was shown in multiple datasets that this system achieved significantly better
DER than the conventional clustering-based speaker diarization system \cite{huang2020speaker,raj2020integration}.
}

\subsubsection{Joint Speech Separation and Diarization}

{\nk
There are also recent researches on the joint modeling of speech separation and
speaker diarization.
Kounades-Bastian et al. \cite{kounades2017algorithm,kounades2017exploiting}
proposed the incorporation of a speech activity model
into speech separation based on the spatial covariance model with non-negative matrix factorization.
They derived the EM algorithm to estimate separated speech and the speech activity of each speaker from the multichannel
overlapped speech. While their method jointly performs speaker diarization
and speech separation, it is based on a statistical modeling,
and the estimation was conducted solely based on the observation, i.e. without 
any model training.

\begin{figure}[t]
  \centering
    \includegraphics[width=\columnwidth]{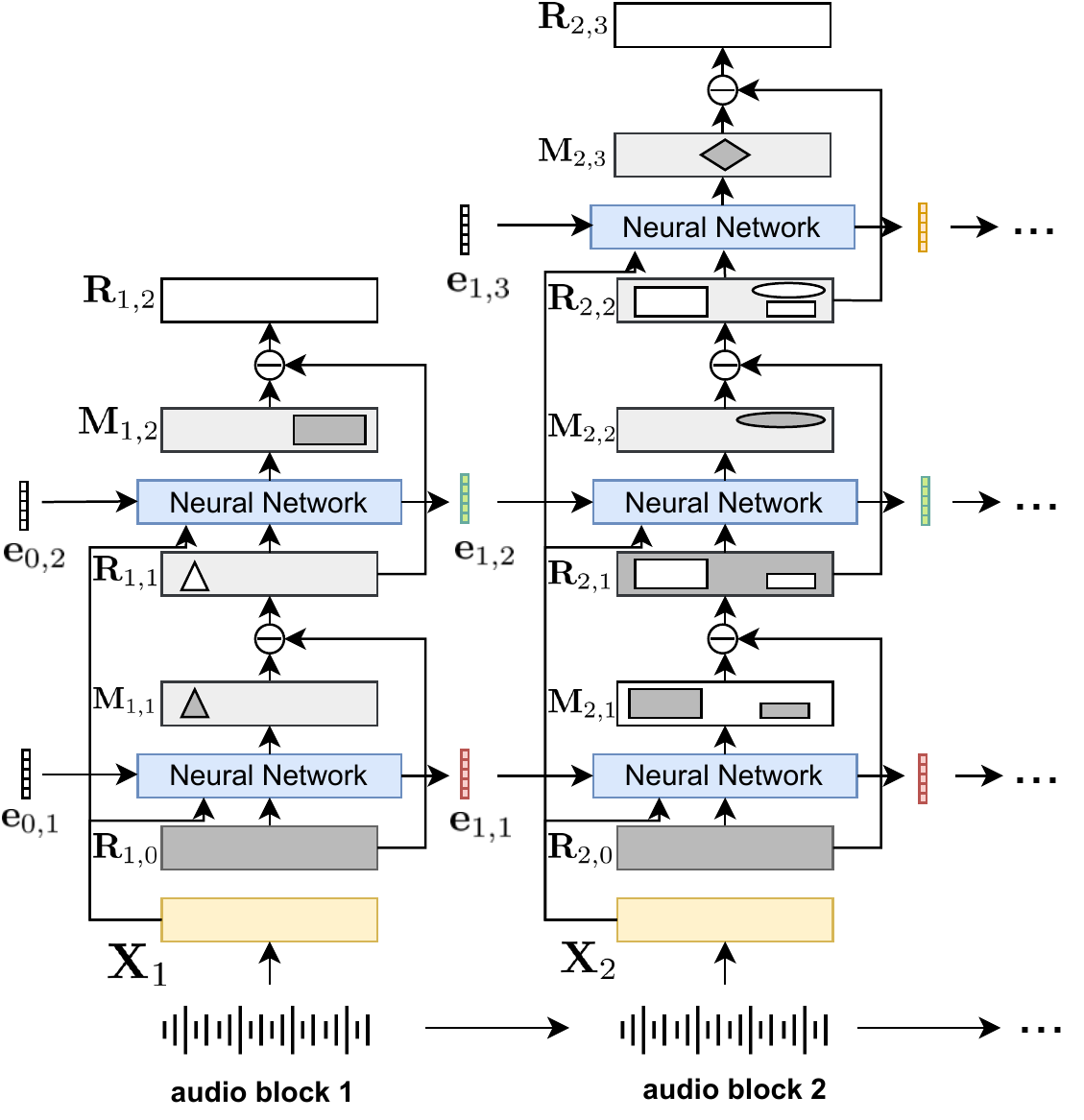}
  \caption{Joint speech separation, speaker counting, and speaker diarization model.}
  \label{fig:neumann}
\end{figure}

{\revision
Neumann et al. \cite{von2019all,kinoshita2020tackling} later proposed 
a trainable model, namely online Recurrent Selective Attention Network (online RSAN), for joint  speech separation,
speaker counting, and speaker diarization based on a single neural network (Fig. \ref{fig:neumann}).
Their neural network takes the input of spectrogram $\mathbf{X}_b\in\mathbb{R}^{T\times F}$, 
 a residual mask $\mathbf{R}_{b,i-1}\in \mathbb{R}^{T\times F}$,
and a speaker embedding $\mathbf{e}_{b-1,i}\in\mathbb{R}^d$, 
where $b$ is the index of the audio block;
$i$, the index of the speaker;
$T$, the length of the audio block;
and $F$, the maximum frequency bin of the spectrogram.
It outputs the speech mask $\mathbf{M}_{b,i}\in\mathbb{R}^{T\times F}$
and an updated speaker embedding
for the speaker corresponding to $\mathbf{e}_{b,i}$. 
The neural network is applied in an iterative fashion for each audio block $b$, and for each speaker $i$ as follows:
\begin{enumerate}
    \item Repeat (a) and (b) for $b=1,2,...$
\begin{enumerate}
\item $\mathbf{R}_{b,0}=\mathbf{1}$
\item Repeat (i)-(iii) for $i=1,2,...$ until being stopped at (iii).
\begin{enumerate}
\item $\mathbf{M}_{b,i}, \mathbf{e}_{b,i} = \mathrm{NN}(\mathbf{X}_b, \mathbf{R}_{b,i-1}, \mathbf{e}_{b-1,i})$ \\ 
{\footnotesize ($\mathbf{e}_{b-1,i}$ is set to $\mathbf{0}$ if it was not calculated previously)} 
\item  $\mathbf{R}_{b,i} = \max(\mathbf{R}_{b,i-1}-\mathbf{M}_{b,i},\mathbf{0})$
\item If $\frac{1}{TF}\sum_{t,f} \mathbf{R}_{b,i}(t,f) < threshold$, stop iteration.
\end{enumerate}
\end{enumerate}
\end{enumerate}

A separated speech for speaker $i$ at audio block $b$ can be obtained by $\mathbf{M}_{b,i}\odot \mathbf{X}_b$ where $\odot$ is the element-wise 
multiplication. 
The speaker embedding $\mathbf{e}_{b,i}$ is used to keep track of the speaker of adjacent blocks.
Thanks to the iterative approach, 
this neural network can cope with the variable number of speakers while
jointly performing speech separation and speaker diarization.
The online RSAN was evaluated by using real meeting dataset with up to six speakers, and 
showed better results than the clustering-based method \cite{kinoshita2020tackling}.}

}

\subsubsection{Fully End-to-end Neural Diarization}
\label{End-to-End Neural Diarization}

\begin{figure}[t]
  \centering
    \includegraphics[width=\columnwidth]{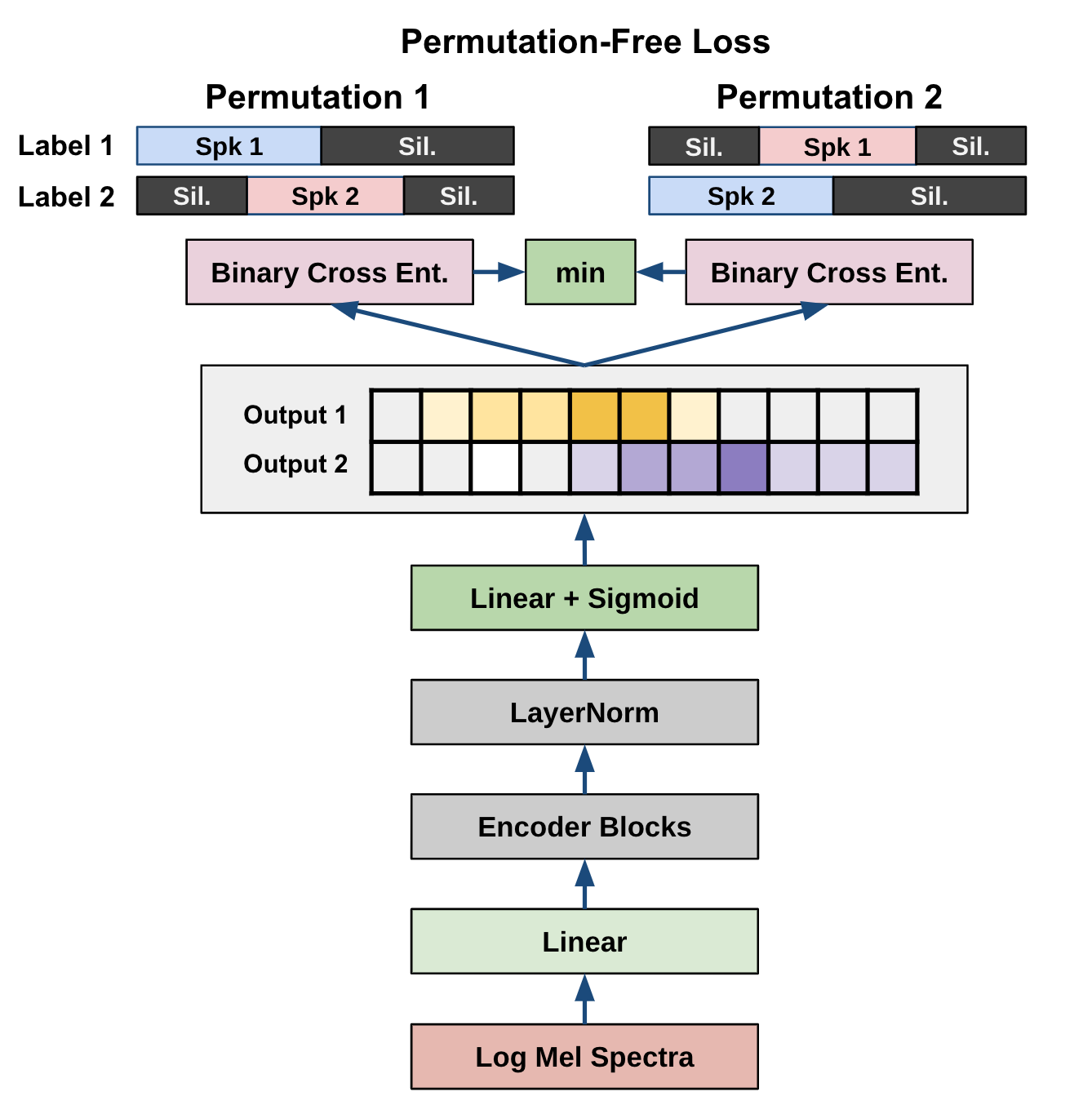}
  \caption{Two-speaker end-to-end neural diarization model}
  \label{fig:eend}
\end{figure}

{\nk 
Recently, the framework called EEND was proposed \cite{fujita2019end,fujita2019end2},
which performs all the speaker diarization procedures 
based on a single neural network.
The architecture of EEND is shown in Fig. \ref{fig:eend}.
An input to the EEND model is a $T$-length sequence of acoustic features (e.g., log Mel-filterbank), $\mathbf{X}=(\mathbf{x}_t\in\mathbb{R}^F|t=1,\ldots,T)$.
A neural network then outputs the corresponding speaker label sequence $\mathbf{Y}=(\mathbf{y}_t|t=1,\ldots,T)$
where $\mathbf{y}_t=[y_{t,k}\in\{0,1\} | k=1,\ldots,K]$. 
Here, $y_{t,k}=1$ represents the speech activity of the speaker $k$ at the time frame $t$, and 
$K$ is the maximum number of speakers that the neural network can output.
Importantly, $y_{t,k}$ and $y_{t,k'}$ can be both 1 for different speakers $k$ and $k'$, indicating that these two speakers $k$ and $k'$ are speaking simultaneously (i.e. overlapping speech).
The neural network is trained to maximize $\log P(\mathbf{Y}|\mathbf{X})\sim\sum_t\sum_k \log P(y_{t,k}|\mathbf{X})$ over the training data
by assuming the conditional independence of the output $y_{t,k}$. 
Because there can be multiple candidates of the reference label $\mathbf{Y}$ by swapping the speaker index $k$, 
the loss function is calculated for all possible reference labels and the reference label that has the minimum loss is
used for the error back-propagation, which is inspired by the permutation free objective used in speech separation \cite{kolbaek2017multitalker}.
EEND was initially proposed using a bidirectional long short-term memory (BLSTM) network \cite{fujita2019end}, 
and  was soon extended to the self-attention-based network \cite{fujita2019end2} by showing the state-of-the-art DER for {\revision two-speaker data such as the two-speaker excerpt from the CALLHOME dataset (LDC2001S97) 
and the dialogue audio in the corpus of Spontaneous Japanese \cite{maekawa2003corpus}.}

EEND has multiple advantages. First, EEND can handle overlapping speech in a sound way.
Second, the network is directly optimized toward the maximization of diarization accuracy, by which we can expect a high accuracy.
Third, it can be retrained by a real data (i.e. not synthetic data) just by feeding a reference diarization label
while it is often not straitforward for the prior works.
However, EEND also has several limitations.
First, the model architecture limits the maximum number of speakers
that the model can cope with.
Second, EEND consists of BLSTM or self-attention based neural networks,
making it difficult to do online processing.
Third, it was empirically suggested that EEND tends to overfit to
the distribution of the training data \cite{fujita2019end}.

\begin{figure}[t]
  \centering
  \includegraphics[width=\columnwidth]{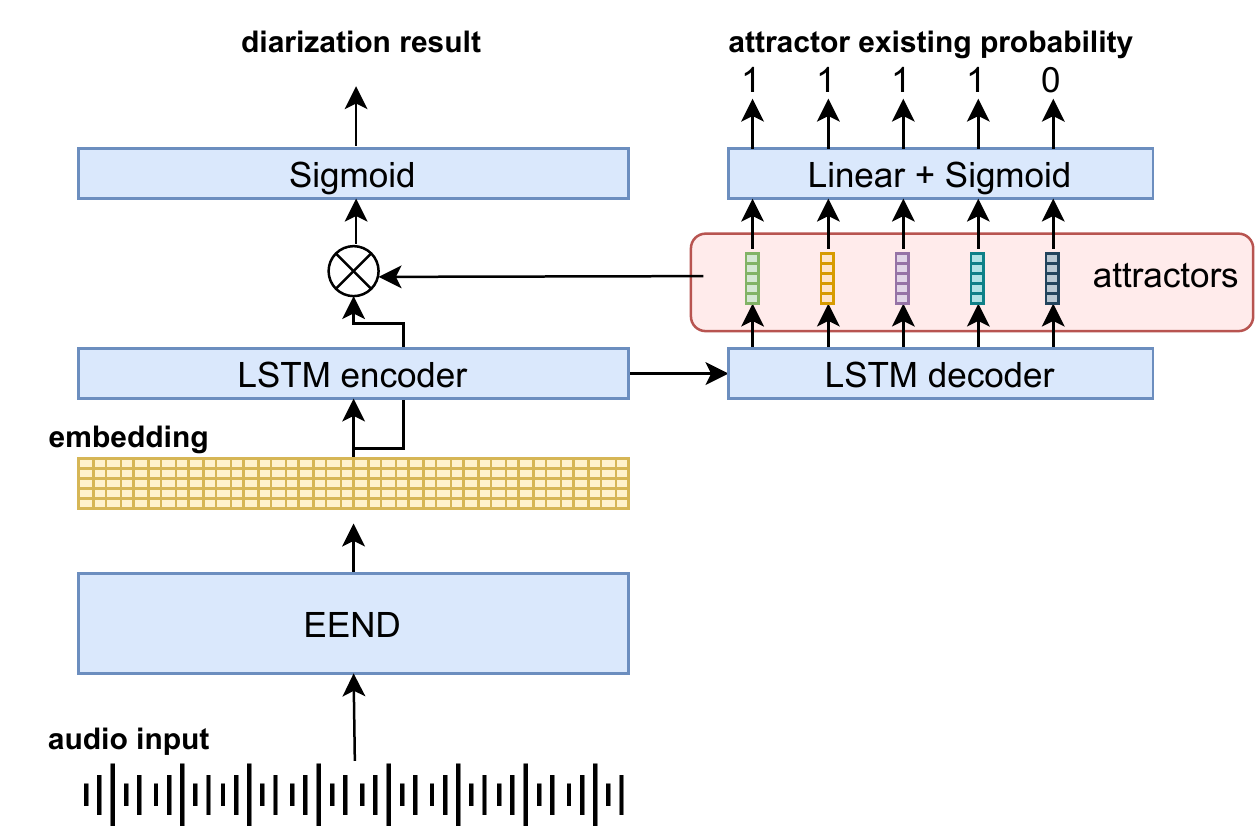}
  \caption{End-to-end neural diarization with encoder-decoder-based attractor (EDA).}
  \label{fig:eda-eend}
\end{figure}

To cope with an unbounded number of speakers,
several extensions of EEND have been investigated.
Horiguchi et al. \cite{horiguchi2020end} proposed 
an extension of EEND with the encoder-decoder-based attractor (EDA)
(Fig. \ref{fig:eda-eend}). 
This method applies an LSTM-based encoder-decoder on
the output of EEND to generate multiple attractors.
Attractors are generated until the attractor existing probability
becomes less than the threshold.
Then, each attractor is multiplied by the embeddings generated
from EEND to calculate the speech activity for each speaker.
{\revision EEND-EDA was evaluated on CALLHOME (two to six speakers) and DIHARD 2 (one to nine speakers) dataset and showed better performance than the clustering-based baseline system.}

On the other hand, Fujita et al. \cite{fujita2020neural} proposed 
another approach to output the speech activity one after another by using a conditional speaker chain rule.
In this method, a neural network
is trained to produce a posterior probability
$P(\mathbf{y}_k|\mathbf{y}_1,\ldots,\mathbf{y}_{k-1},\mathbf{X})$, 
where $\mathbf{y}_k=(y_{t,k}\in \{0,1\}|t=1,\ldots,T)$ is the speech activity for
the $k$-th speaker.
Then, the joint speech activity probability of all speakers can be
estimated from the following speaker-wise conditional chain rule as:
\begin{align}
P(\mathbf{y}_1,\ldots,\mathbf{y}_K|\mathbf{X}) &= \prod_{k=1}^K P(\mathbf{y}_k|\mathbf{y}_1,\ldots,\mathbf{y}_{k-1},\mathbf{X}).
\end{align}
During inference, the neural network is repeatedly applied
until the speech activity $y_k$ for the last estimated speaker approaches zero. 
Kinoshita et al. \cite{kinoshita2020integrating}
proposed a different approach 
that combines EEND and speaker clustering.
In their method, a neural network is trained
to generate speaker embeddings and
the speech activity probability.
Speaker clustering constrained by the estimated speech activity by EEND
is applied to align the estimated speakers among the different processing blocks.

There are also a few recent trials to extend the EEND for online processing.
Xue et al. \cite{xue2020online} proposed
a method using a speaker tracing buffer to
better align the speaker labels of adjacent
processing blocks.
Han et al. \cite{han2020bw} proposed a block 
online version of EEND-EDA \cite{horiguchi2020end} by carrying
the hidden state of the LSTM-encoder to generate the attractors 
block by block.
}


\section{Speaker Diarization in the Context of ASR}
\label{Speaker Diarization in the context of Automatic Speech Recognition}

{\kh From a conventional perspective, speaker diarization is considered a pre-processing step for ASR. In the traditional system structures for speaker diarization, presented in Fig.~\ref{fig:modular}, speech inputs are processed sequentially across the diarization components without considering the ASR performance, which is usually measured using the word error rate (WER). 
{\revision WER is the number of misrecognized words (substitution error, insertion error, and deletion error) divided by the number of reference words.}
One issue is that the tight boundaries of speech segments as the outcomes of speaker diarization have a high chance of causing unexpected word truncation or deletion errors in ASR decoding. In this section we discuss how the speaker diarization systems have been developed in the context of ASR, not only resulting in better WER by preventing speaker diarization from affecting the ASR performance, but also benefiting from ASR artifacts to enhance diarization performance. More recently, there have been a few pioneering proposals made for the joint modeling of speaker diarization and ASR, which will also be introduced in this section.} 

\subsection{Early Works} 
\label{subsec:early-work}
{\tp The lexical information from the ASR output has been employed for the speaker diarization system in a few different ways. First, the earliest approach was the RT03 evaluation \cite{tranter2003investigation} which used word boundary information for the purpose of segmentation. In \cite{tranter2003investigation}, a general ASR system for broadcast news data was built, in which the basic components are segmentation, speaker clustering, speaker adaptation and system combination after ASR decoding from the two sub-systems with the different adaptation methods. {\rrevision The authors used the word boundary information from the ASR system for speech segmentation, and compared it with the BIC-based speech segmentation.  While the performance gain by the ASR-based segmentation was insignificant, this was the first attempt to take advantage of ASR output to enhance the diarization performance.} In addition, the ASR result was used to refine SAD in IBM's submission \cite{huang2007ibm} for RT07 evaluation. The system that appeared in \cite{huang2007ibm} incorporates word alignments from the speaker independent ASR module and refines the SAD result to reduce false alarms so that the speaker diarization system can have better clustering quality. The segmentation system in \cite{silovsky2012incorporation} also takes advantage of word alignments from ASR. The authors in \cite{silovsky2012incorporation} focused on the word-breakage problem, in which the words from the ASR output are truncated by segmentation results since the segmentation results and the decoded word sequences are not aligned. Therefore, word-breakage ratio was proposed to measure the rate of change points detected inside intervals corresponding to words. The DER and word-breakage ratio were used to measure the influence of the word truncation problem. 
While the aforementioned early works of speaker diarization systems that leverage the ASR output focus on the word alignment information to refine the SAD or segmentation result, the speaker diarization system in \cite{canseco2004speaker} created a dictionary for the phrases commonly appearing in broadcast news. The phrases in this dictionary provide the identity of who is speaking, who will speak and who spoke in the broadcast news scenario. For example, ``This is [name]" indicates who was the speaker of the broadcast news section. Although the early studies on speaker diarization did not fully leverage the lexical information to drastically improve the DER, the idea of integrating the information from ASR output has been adopted by many studies to refine or improve the speaker diarization output.
}

\begin{figure}[t]
  \centering
    \includegraphics[width=\columnwidth]{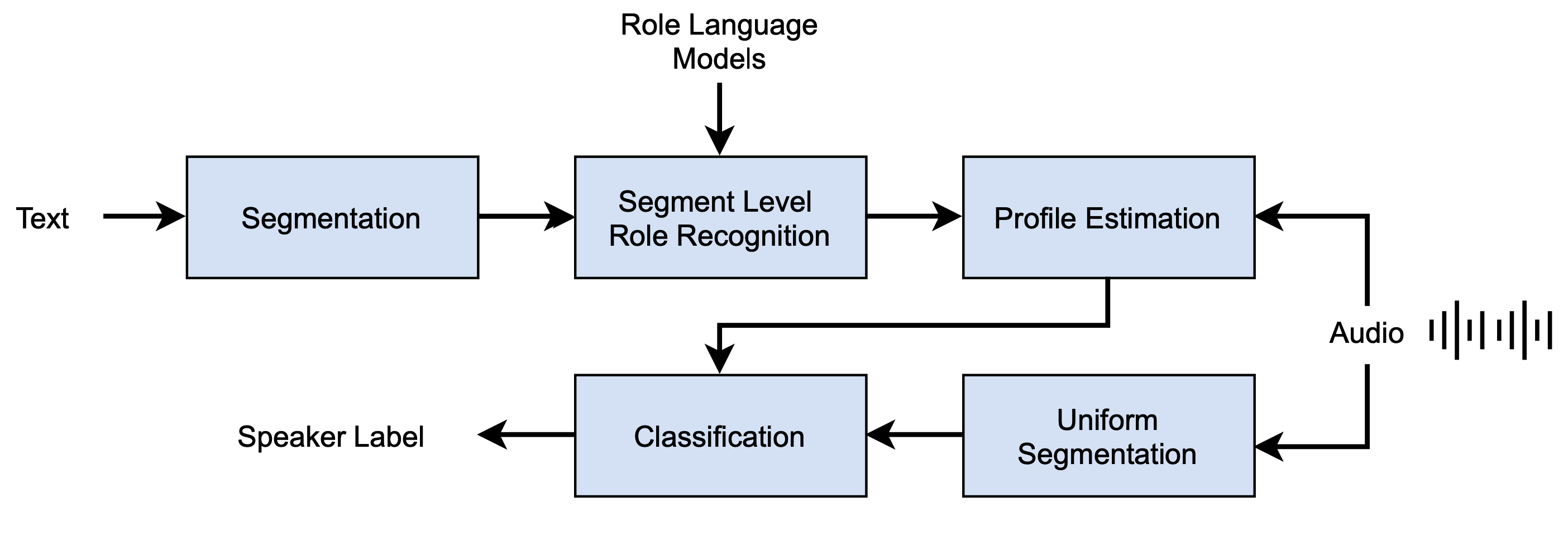}
  \caption{Integration of lexical information and acoustic information.}
  \label{fig:ling_aided_diar}
\end{figure}

\begin{figure}[t]
  \centering
    \includegraphics[width=\columnwidth]{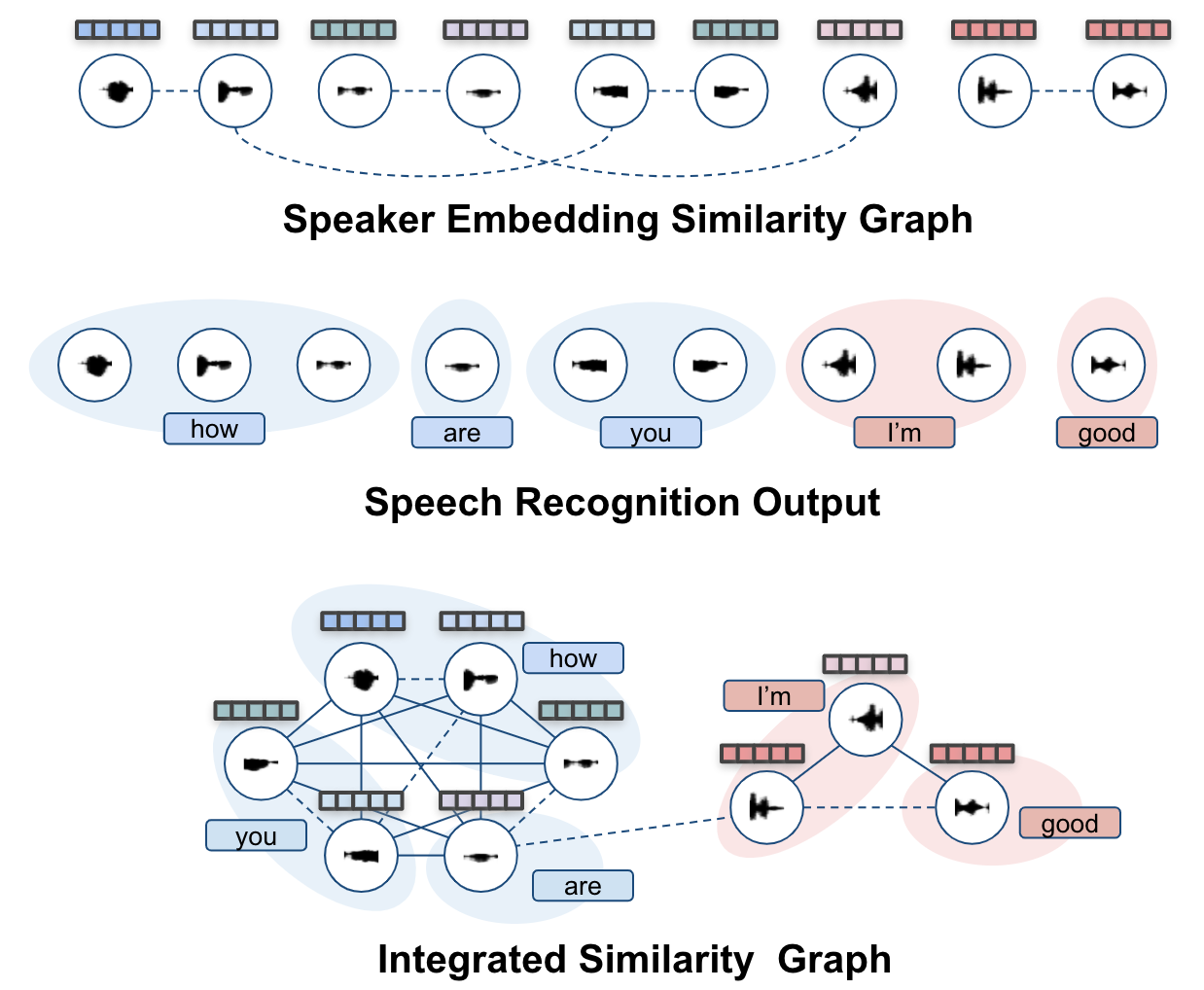}
  \caption{Integration of lexical information and acoustic information.}
  \label{fig:integration-of-lexical-info}
\end{figure}

\subsection{Using lexical information from ASR} 
{\tp
The more recent speaker diarization systems that take advantage of the ASR transcript have employed a DNN model to capture the linguistic pattern in the given ASR output to enhance the speaker diarization result. The authors in \cite{flemotomos2019linguistically} proposed a way of using the linguistic information for the speaker diarization task where participants have distinct roles that are known to the speaker diarization system. Fig.~\ref{fig:ling_aided_diar} shows the diagram of the speaker diarization system discussed in \cite{flemotomos2019linguistically}. In this system, a neural text-based speaker change detector and a text-based role recognizer are employed. By using both linguistic and acoustic information, DER was significantly improved compared with the acoustic only system. 

Lexical information from the ASR output was also used for speaker segmentation \cite{park2018multimodal} by employing a sequence-to-sequence model that outputs speaker turn tokens. Based on the estimated speaker turn, the input utterance is segmented accordingly. The experimental results in \cite{park2018multimodal} indicate that using both acoustic and lexical information can be exploited and an extra advantage can be obtained owing to the word boundaries we get from the ASR output.

The authors of \cite{park2019speaker} presented follow-up research within the above thread. Unlike the system in \cite{park2018multimodal}, the lexical information from the ASR module was integrated with the speech segment clustering process by employing an integrated adjacency matrix. The adjacency matrix is obtained from the max operation between the acoustic information created from affinities among audio segments and lexical information matrix created by segmenting the word sequence into word chunks that are likely to be spoken by the same speaker.
Fig.~\ref{fig:integration-of-lexical-info} presents a diagram that explains how lexical information is integrated in an affinity matrix with acoustic information. The integrated adjacency matrix leads to an improved speaker diarization performance for the CALLHOME American English dataset.

}

\label{Using lexical information from ASR}

\subsection{Joint ASR and Speaker Diarization with Deep Learning} 
\label{Joint ASR and SD with deep learning}

{\nk
Motivated by the recent success of deep learning and end-to-end modeling,
several models have been proposed
to jointly perform ASR and speaker diarization. 
As with the previous section, the ASR results contain a strong cue to improve speaker diarization. On the other hand, speaker diarization results can be used to improve the accuracy of ASR, for example, by adapting the ASR model toward each estimated speaker. Joint modeling can leverage such inter-dependency to improve both ASR and speaker diarization. 
In the evaluation,
{\revision a WER metric that counts word hypotheses with speaker-attribution errors as misrecognized words}, such as
speaker-attributed WER \cite{FiscusEtAl:rt07} or concatenated minimum-permutation WER (cpWER) \cite{watanabe2020chime},
is often used.
ASR-specific metrics (e.g., speaker-agnostic WER) or 
diarization-specific metrics (e.g., DER {\rrevision mentioned in Section \ref{subsubsec:DER}}) are also used complementarily.

\begin{figure}[t]
  \centering
    \includegraphics[width=\columnwidth]{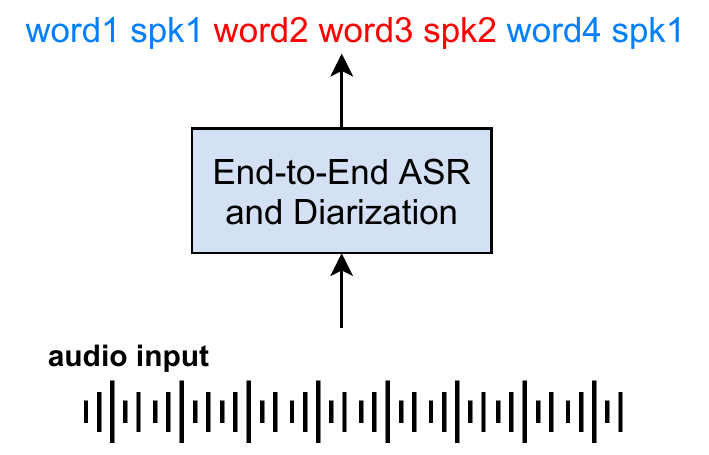}
  \caption{Joint ASR and diarization by inserting a speaker tag in the transcription.}
  \label{fig:asr-sd-1}
\end{figure}


The first approach is the introduction of a speaker tag in the transcription of end-to-end ASR models (Fig. \ref{fig:asr-sd-1}).
Shafey et al. \cite{Shafey2019} proposed to 
insert a speaker role tag (e.g., $\langle$doctor$\rangle$ 
and $\langle$patient$\rangle$)
into the output of a recurrent neural network-transducer (RNN-T)-based ASR system.
{\revision This method was evaluated by using doctor-patient conversation,
and a significant reduction in WDER was reported with 
a marginal degradation of WER.}
Similarly, Mao et al. \cite{mao2020speech} proposed the insertion of a speaker identity tag
into the output of
an attention-based encoder-decoder ASR system,
{\revision and showed an improvement of DER especially when the oracle utterance boundaries were not given.
The works by Shafey et al. and Mao et al. showed that the insertion of speaker tags is a  simple and promising way to jointly perform ASR and speaker diarization.} 
On the other hand, the speaker roles or speaker identity tags need to be determined and fixed during training. Thus, it is difficult to cope with an arbitrary number of speakers using this approach.

\begin{figure}[t]
  \centering
    \includegraphics[width=\columnwidth]{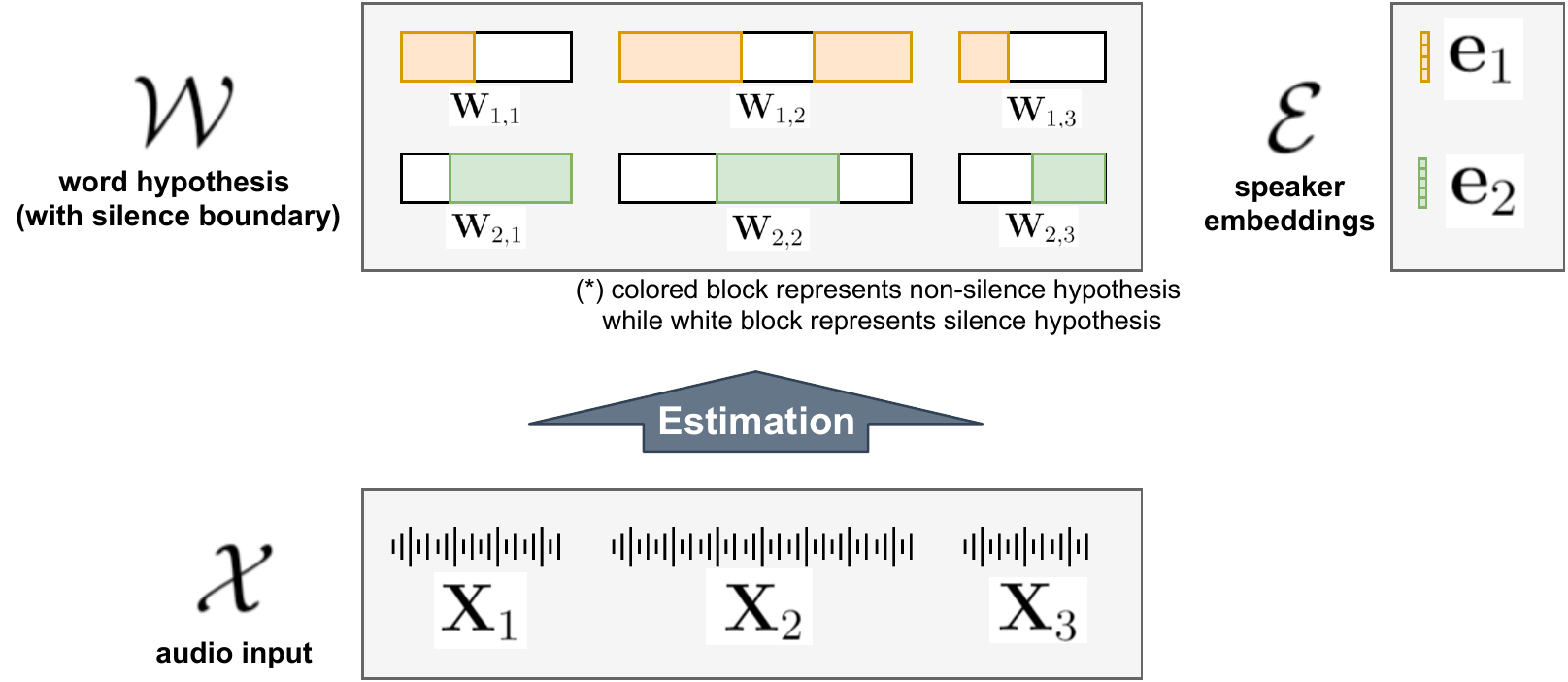}
  \caption{Joint decoding framework for ASR and speaker diarization.}
  \label{fig:asr-sd-3}
\end{figure}

The second approach is a 
MAP-based joint decoding framework.
Kanda et al. \cite{kanda2019simultaneous} formulated 
the joint decoding of ASR and speaker diarization as follows (see also Fig. \ref{fig:asr-sd-3}).
Assume that a sequence of observations is represented by
$\mathcal{X} = \{\mathbf{X}_1,\ldots, \mathbf{X}_U\}$, 
where $U$ denotes the number of segments (e.g., generated by applying {\rrevision SAD} on a long audio) 
and $\mathbf{X}_u$ denotes the acoustic feature sequence
of the $u$-th segment.
Further assume that
word hypotheses with time boundary information are represented by
 $\mathcal{W}=\{\mathbf{W}_1,\ldots,\mathbf{W}_U\}$
 where $\mathbf{W}_u$ is the speech recognition hypothesis 
 corresponding to the segment $u$. 
 Here,
 $\mathbf{W}_u = (\mathbf{W}_{1,u}, ..., \mathbf{W}_{K,u})$ 
 contains all the speakers' hypotheses in the segment $u$ 
where $K$ denotes the number of speakers, and $\mathbf{W}_{k,u}$ 
represents the speech recognition hypothesis of the $k$-th speaker 
of the segment $u$.
Finally, a tuple of speaker embeddings $\mathcal{E}=(\mathbf{e}_1,\ldots,\mathbf{e}_K)$,
where $\mathbf{e}_j\in\mathbb{R}^d$ is the $d$-dimensional speaker embedding of the $k$-th speaker,
is also assumed. With all these notations, the joint decoding framework of multispeaker
ASR and diarization can be formulated as a problem to find most likely $\hat{\mathcal{W}}$ as follows:
\begin{align}
  \hat{\mathcal{W}} =&\argmax_{\mathcal{W}} P(\mathcal{W}|\mathcal{X})\\
  =&\argmax_{\mathcal{W}}\{\sum_{\mathcal{E}}P(\mathcal{W},\mathcal{E}|\mathcal{X})\} \\
  \approx&\argmax_{\mathcal{W}}\{\max_{\mathcal{E}}P(\mathcal{W},\mathcal{E}|\mathcal{X})\},
\end{align}
where the Viterbi approximation is applied to obtain the final equation.
This maximization problem is further decomposed into two iterative problems as follows:
\begin{align}
\hat{\mathcal{W}}^{(i)}=& \argmax_{\mathcal{W}}
P(\mathcal{W}|\hat{\mathcal{E}}^{(i-1)},\mathcal{X}),\label{eq:joint1}  \\
\hat{\mathcal{E}}^{(i)}=& \argmax_{\mathcal{E}} 
P(\mathcal{E}|\hat{\mathcal{W}}^{(i)},\mathcal{X}), \label{eq:joint2}
\end{align}
where $i$ is the iteration index of the procedure.
In \cite{kanda2019simultaneous},
Eq. \eqref{eq:joint1} is modeled by the target speaker ASR \cite{zmolikova2017speaker,delcroix2018single,delcroix2019end,kanda2019auxiliary}
and Eq. \eqref{eq:joint2} is modeled by the overlap-aware speaker embedding
estimation.
This method obtains a speaker-attributed WER similar to 
that of the target-speaker ASR with oracle speaker embeddings
{\revision for two-speaker conversation data of the Corpus of Spontaneous Japanese \cite{maekawa2003corpus}}.
On the other hand, it requires an iterative application of the target-speaker ASR
and a speaker embedding extraction scheme, which make it challenging to 
apply the method in online mode.


\begin{figure}[t]
  \centering
  \includegraphics[width=\columnwidth]{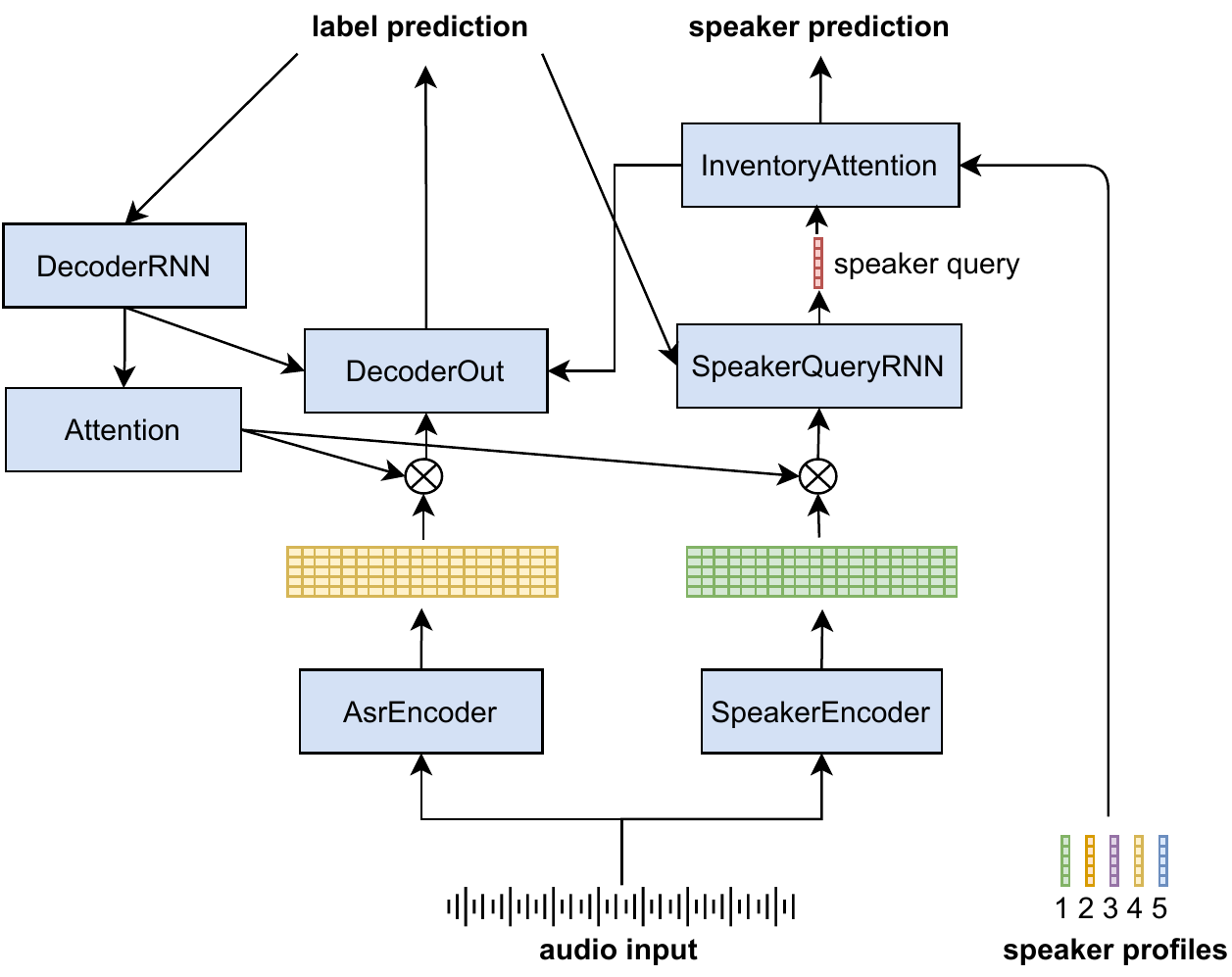}
  \caption{End-to-end speaker-attributed ASR}
  \label{fig:e2e-sa-asr}
\end{figure}

As a third line of approaches,     
end-to-end speaker-attributed ASR (SA-ASR) model was recently proposed 
to jointly perform speaker counting, multi-talker ASR, and speaker identification \cite{kanda2020joint,kanda2020minimum}.
Contrary to the first two approaches, 
the end-to-end SA-ASR model takes the additional input of speaker profiles
and identifies the index of speaker profiles 
based on the attention mechanism (Fig.~\ref{fig:e2e-sa-asr}).
Thanks to the attention mechanism 
for the speaker identification and 
multi-talker ASR capability based on 
serialized output training \cite{kanda2020sot},
there is no limitation in the number of
speakers that the model can cope with. 
In case relevant speaker profiles are supplied
in the inference,
the end-to-end SA-ASR model can automatically
transcribe the utterance
while identifying the speaker of each utterance based on the supplied 
speaker profiles.
On the other hand, in case of the relevant speaker profiles 
cannot be used prior to the inference,
the end-to-end SA-ASR model can still be applied using dummy profiles,
and the speaker clustering on the internal speaker 
embeddings of
the end-to-end SA-ASR model (``speaker query" in Fig. \ref{fig:e2e-sa-asr}) 
is used to 
diarize the speaker \cite{kanda2021investigation}.
{\revision The end-to-end SA-ASR model was evaluated by using the LibriCSS dataset \cite{chen2020continuous}, and exhibited significantly better cpWER than the combination of multitalker ASR and speaker diarization \cite{kanda2021end}.} 
}


\section{Diarization Evaluation Series and Datasets}
\label{Diarization Evaluation Series and datasets}

{\nk
{\revision
This section describes the evaluation series and the commonly used datasets for speaker diarization evaluations. The summary of the most commonly used datasets that include English is shown in Table \ref{tab:datasets}.
}
}

{\revision 
\begin{table*}[t]
\centering
\caption{\revision Diarization Evaluation Datasets}
\vspace{0.5ex}
\label{tab:datasets}
{

\begin{tabular}{r|l|l|l|l} \hline
             & Language & Size (hr) & Style        & \# Spkr.  \\
             \hline
             \hline
CALLHOME     & Multilingual & 20          & Conversation & 2--7        \\
AMI          & English & 100          & Meeting      & 3--5        \\
ICSI meeting & English & 72          & Meeting      & 3--10       \\
CHiME-5/6    & English & 50          & Conversation & 4  \\
VoxConverse  & Multilingual$^\dagger$ & 74          & YouTube video & 1--21\\
LibriCSS     & English & 10          & Read speech  & 8\\ \hline
DH I Tr.1,2 & Multilingual & 19(dev), 21(eval) & Miscellaneous &  1--7 \\
DH II Tr.1,2 & Multilingual & 24(dev), 22(eval)& Miscellaneous &  1--8 \\
DH II Tr.3,4 & Multilingual &  262(dev), 31(eval) & Miscellaneous & 4 \\
DH III Tr.1,2 & Multilingual &  34(dev), 33(eval)     & Miscellaneous & 1--7 \\ \hline
\end{tabular}
}
{\footnotesize
\\$^\dagger$ Most of the contents are English while there are few non-English contents.
}
\end{table*}

}

\begin{itemize}

\item \textbf{CALLHOME: NIST SRE 2000 (LDC2001S97)}\\
NIST SRE 2000 (Disk-8), often referred to as the CALLHOME dataset, is the most widely used dataset for speaker diarization in 
recent papers. This dataset contains 500 sessions of multilingual telephonic speech. Each session has two to seven speakers with two dominant speakers in each conversation. 

\item \textbf{AMI Corpus}\\
The AMI database \cite{carletta2005ami} includes 100 h of meeting recordings from multiple sites in 171 meeting sessions. The AMI database provides an audio source recorded using lapel microphones that are separately recorded and amplified for each speaker. Another audio source is recorded using omnidirectional microphone arrays mounted on the table while meeting. The AMI database is a suitable dataset for the evaluation of speaker diarization systems integrated with the ASR module since AMI provides forced alignment data that contains word and phoneme level timings along with the transcript and speaker label. Each meeting session has three to five speakers. 
\item \textbf{ICSI Meeting Corpus}\\
The ICSI meeting corpus \cite{Janin03} contains 75 meeting corpus with four meeting types. The ICSI meeting corpus provides word level timing along with the transcript and speaker label. The audio source is recorded using close-talking individual microphone and six tabletop microphones to provide speaker-specific channel and multichannel recording. Each meeting has 3 to 10 participants. 

\item \textbf{CHiME-5/6 challenge and its dataset}\\
{\revision
The CHiME-5 challenge \cite{barker2018fifth} and CHiME-6 challenge \cite{watanabe2020chime} were designed as
series of ASR competitions for the daily conversation of multiple speakers.
The dataset was provided at the CHiME-5 challenge, 
and it contains 50 h of multiparty real conversations in the everyday home environment.
It contains speaker labels, segmentation, and corresponding transcriptions.
The audio source is recorded using six four-channel microphone arrays located in the kitchen and dining/living rooms in a house and also binaural microphones worn by participants.
The number of participants is fixed at four.
 While the oracle diarization results were allowed to be used for the ASR task in the CHiME-5 challenge, 
CHiME-6 challenge track 2 requires the result of both ASR and diarization.
The primary evaluation metric for such a track was cpWER, which counts both the speaker-attributed 
errors and word recognition errors in the WER calculation.
DER and JER were also evaluated as secondary metrics without ``score collar'' and with  overlapped regions. 
The CHiME-5/6 corpus was also used as one track in the DIHARD 2 challenge.}\\

\item \textbf{VoxSRC Challenge and VoxConverse corpus}\\
{\tp {\revision
The VoxCeleb Speaker Recognition Challenge (VoxSRC) is the recent evaluation series for speaker recognition systems \cite{chung2019voxsrc,nagrani2020voxsrc}.
The goal of VoxSRC is to test how well the current technology can cope with the speech ``in the wild".
This evaluation series initially started with a pure speaker verification task \cite{chung2019voxsrc},
and the diarization task was added as track 4 at the latest evaluation at the VoxCeleb Speaker Recognition Challenge 2020 (VoxSRC-20) \cite{nagrani2020voxsrc}.}
The VoxConverse dataset \cite{chung2020spot} was used for the speaker diarization task with DER as the primary metric, and JER as the secondary metric.
The VoxConverse dataset contains 74 h of human conversation extracted from YouTube videos.
The dataset is divided into development set (20.3 h, 216 recordings),
and test set (53.5 h, 310 recordings).
The number of speakers in each recording
has a wide range of variety from
1 speaker to 21 speakers.
The audio includes
various types of noises such as background music, laughter, etc.
It also contains a significant portion of overlapping speech 
from 0\% to 30.1\% depending on the recording.
While the dataset contains the visual information as well as audio,
as of June 2021,
only the audio of the development set was released under 
a Creative Commons Attribution 4.0 International License
for research purposes. 
The audio of the evaluation set was used 
as a blind test set.
}

\item \textbf{LibriCSS} \\
{\nk 
The LibriCSS corpus \cite{chen2020continuous} contains
10 h of multichannel recordings and was designed for the research of speech separation, speech recognition,
and speaker diarization.
It was created by playing the audio in the LibriSpeech corpus \cite{panayotov2015librispeech} in 
a real meeting room, and recorded using
a 7-channel microphone array.
It 
consists of 10 sessions,
each of which is further decomposed to 
six 10-min mini-sessions.
Each mini-session was made by audio of eight speakers
and designed to have 
different overlap ratios from 0\% to 40\%.
To facilitate the research, the baseline system for speech separation and 
ASR \cite{chen2020continuous} and the baseline system that integrates speech separation, speaker diarization
and ASR \cite{raj2020integration} have been developed and released.
}

\item \textbf{DIHARD Challenge and its dataset}\\
{\revision 
DIHARD evaluation  \cite{ryant2018first, ryant2019second} focuses on the performance gap of state-of-the-art diarization systems on challenging domains. The first DIHARD challenge, DIHARD 1, started with track 1 (oracle SAD) and 
track 2 (system SAD). The evaluation data was a collection of various corpus.
It includes very challenging datasets such as clinical interviews, web videos, and speech in the wild (e.g., recordings in restaurants), as well as relatively less challenging datasets, such as CTS and audio books to diversify the domains.
DIHARD 2 additionally included multichannel speaker diarization task in track 3 (oracle SAD) and track 4 (system SAD) using the recordings from the CHiME-5 corpus \cite{barker2018fifth}. 
In the latest DIHARD challenge, DIHARD 3, 
the CTS dataset was added, whereas multichannel tracks 3 and 4 were excluded.
The DIHARD challenge employs DER and JER for the evaluation metric without ``score collar'' and with overlap region.}

\item \textbf{Rich Transcription Evaluation Series}\\
{\kh 
The RT Evaluation \cite{nistrtwebsite} is the pioneering evaluation series of initiating deeper investigation on speaker diarization in relation to ASR. The main purpose of this effort was to create ASR technologies that would produce transcriptions with descriptive metadata, like who said when, where speaker diarization plays in. Thus, the main tasks in the evaluation were ASR and speaker diarization. The domains of the data of interest were broadcast news, CTS and meeting recordings with multiple participants. Throughout the period of 2002-2009, the RT evaluation series promoted and gauged advances in speaker diarization and ASR technology.} 
{\tp {\revision The evaluations among this period are named as RT evaluations (RT-02, RT-03S, RT-03F, and RT-05F) and RT Meeting Recognition (RT-06S, RT-07S and RT-09). These evaluations and their datasets include speaker diarization evaluation as a part of automatic metadata extraction (MDE).}}

\item \textbf{Other datasets}

{\revision There are also several corpora that have been used for the diarization research but not covered in the list above.
The Corpus of Spontaneous Japanese \cite{maekawa2003corpus} contains about 12 h of two-speaker dialogue recorded using headset microphones. AISHELL-4 \cite{fu2021aishell} is a relatively new Mandarin Chinese dataset containing 118 h of four to eight speakers in a conference scenario. It is recorded by 8-ch circular microphone array as well as headset microphones for each participant. The ESTER-1~\cite{gravier2004ester} and ESTER-2~\cite{galliano2009ester} evaluation campaign datasets are a set of French recordings designed for three task category: Segmentation (S), Transcription (T) and Information Extraction (E). In the ESTER-1 and ESTER-2 evaluation campaign, speaker diarization was evaluated as one of the core tasks among other tasks including speaker tracking, sound event tracking, and transcriptions. The datasets for ESTER-1 and ESTER-2 include 100 h and 150 h of manually transcribed French radio broadcast news, respectively. 
ETAPE~\cite{gravier2012etape} is also a French speech processing evaluation dataset that contains 36 h of TV and radio shows with both prepared and spontaneous speech. Unlike the ESTER evaluation series, ETAPE targets cross-show speaker diarization.  
}

\end{itemize}

\label{Perspectives}
\section{Applications}
\subsection{Meeting Transcription}
{\dd 

The goal of meeting transcription is to automatically generate speaker-attributed transcripts during real-life meetings based on their audio and optional video recordings. Accurate meeting transcription is among the processing steps in a pipeline for several tasks, such as, summarization and topic extraction. Similarly, the same transcription system can be used in other domains such as healthcare~\cite{Chiu18}. 

Although this task was introduced by NIST in the RT evaluation series back in 2003~\cite{FiscusEtAl:rt07,Janin03,Carletta06}, the initial systems had very poor performance, and consequently commercialization of the technology was not possible. However, recent advances in the areas of speech recognition~\cite{Xiong16, Saon17}, far-field speech processing~\cite{Yoshioka15b, Du16, Li17}, speaker ID and diarization~\cite{Dimitriadis17,zhang2019fully,sell2018diarization}, have greatly improved the speaker-attributed transcription accuracy, enabling such commercialization.  Bimodal processing combining cameras with  microphone arrays has further improved the overall performance~\cite{He16,HeGDG17}.  

The variety of application scenarios, customer needs, and business scope, different constraints may be imposed on meeting transcription systems. For example, it is most often required to provide the resulting transcriptions in low latency, making the diarization and recognition even more challenging. However, the architecture of the transcription system can substantially improve the overall performance, e.g., by using microphone arrays of known geometry as the input device. Also, in the case where the expected meeting attendees are known beforehand, the transcription system can further improve speaker attribution, all while  providing the exact name of the speaker, instead of randomly generated discrete speaker labels.  

Two different scenarios in this space are presented: first, a fixed-geometry microphone array combined with a fish-eye camera system. Second, an ad-hoc geometry microphone array system without a camera. In both scenarios, a ``non-binding'' list of participants and their corresponding speaker profiles are considered to be known. In particular, the transcription system has access to the invitees' names and profiles; however, the actual attendees may not accurately match those invited. As such, there is an option to include ``unannounced'' participants. In addition, some of the invitees may not have profiles. In both scenarios, there is a constraint of low-latency transcriptions, where initial results need to be shown with low latency. The finalized  results can be updated later in an offline mode. Some of the technical challenges to overcome are~\cite{yoshioka2019advances}:
\begin{enumerate}
    \item Although ASR on overlapping speech is one of the main challenges in meeting transcription, limited progress has been made over the years. Numerous multichannel speech separation methods have been proposed based on independent component analysis (ICA) or spatial clustering~\cite{Buchner05,Sawada07,Nesta11,Sawada11,Ito14,Drude17}, but their application to a meeting setup had limited success. In addition, neural network-based separation methods such as permutation invariant training (PIT)~\cite{kolbaek2017multitalker} or deep clustering (DC)~\cite{hershey2016deep} cannot adequately address  reverberation and background noise~\cite{Maciejewski18}. 

    \item Flexible framework: It is desirable that the transcription system is capable of processing all the available information, such as the multichannel audio and visual cues. The system needs to process a dynamically changing number of audio channels without loss of performance. As such, the architecture needs to be modular enough to encompass the different settings.
 
    \item The speaker-attributed ASR of natural meetings requires online or streaming ASR, audio pre-processing such as dereverberation, and accurate diarization and speaker identification. These multiple processing steps are usually optimized separately and thus, the overall pipeline is most frequently inefficient. 
    
    \item The use of multiple, unsynchronized audio streams, e.g., audio capturing using mobile devices, adds complexity to the meeting setup and processing. In return, we gain a potentially better spatial coverage since the devices are usually  distributed around the room and near the speakers. As part of the application scenario, the meeting participants bring their personal devices, which can be repurposed to improve the overall quality of meeting transcription quality. On the other hand, while there are several pioneering studies~\cite{Araki18}, it is unclear what the best strategies are for consolidating multiple asynchronous audio streams and to what extent they work for natural meetings in online and offline setups. 
\end{enumerate}

Based on these considerations, 
an architecture of a meeting transcription system with asynchronous distant microphones has been proposed in ~\cite{yoshioka2019meeting}. In this work, various fusion strategies have been investigated: from early fusion beamforming of the audio signals, to mid-fusion combination of senones per channel,  to late fusion combination of the diarization and ASR results~\cite{stolcke2019dover}. The resulting system performance was benchmarked  on real-world meeting recordings against fixed geometry systems. As aforementioned, the requirement of speaker-attributed transcriptions with low latency was also adhered to. In addition to the end-to-end system analysis, the paper~\cite{yoshioka2019meeting} proposed the idea of ``leave-one-out beamforming'' in the asynchronous multi-microphone setup, enriching the ``diversity''  of the resulting signals, as proposed in~\cite{Stolcke11}. Finally, it is described how an online, incremental version of recognizer output voting error reduction (ROVER) \cite{fiscus1997post} can process both the ASR and diarization outputs, enhancing the overall speaker-attributed ASR performance.

}

\subsection{Conversational Interaction Analysis and Behavioral Modeling}
{\sn
{\revision Speech and spoken language are central to conversational interactions. They carry crucial information about a speaker’s intent, emotions, identity, age and other individual and interpersonal  traits and state variables including health state. Computational advances are increasingly allowing access to such rich information \cite{narayanan2013behavioral,Bone2017SignalProcessingandMachine}.} 
For example, knowing how much, and how, a child speaks in an interaction contains critical information about the child's developmental state, and offers clues to clinicians in diagnosing disorders such as autism \cite{Kumar2020SpeakerDiarizationforNaturalistic}. Such analyses are made possible by capturing and processing the audio recordings of the interactions, which often involve two or more people. An important foundational step is the identification and association of the speech portions belonging to specific individuals involved in the conversation.  The technologies providing these capabilities are SAD and speaker diarization. Speech portions segmented with speaker-specific information provided by speaker diarization, by itself without any explicit lexical transcription, can offer important information to domain experts who can take advantage of speaker diarization results for quantitative turn-taking analysis.  

A domain that is the most relevant in such analyses of spoken conversational interactions relates to behavioral signal processing (BSP) \cite{georgiou2011behavioral, narayanan2013behavioral}, which refers to the technology and algorithms for modeling and understanding human communicative, affective and social behaviors. For example, these may include analyzing how positive or negative a person is, how empathic an individual {\revision is} toward another, what the behavior patterns reveal about the relationship status, and the health condition of an individual \cite{Bone2017SignalProcessingandMachine}. BSP involves addressing all the complexities of spontaneous interactions in conversations with additional challenges involved in handling and understanding emotional, social and interpersonal behavioral dynamics revealed through verbal and nonverbal cues of the interaction participants. Therefore, the knowledge of speaker specific vocal information plays a significant role in BSP, requiring highly accurate speaker diarization performance. For example, the speaker diarization module is employed as a pre-processing module for analyzing psychotherapy mechanisms and quality \cite{xiao2016technology} and suicide risk assessment \cite{chakravarthula2020automatic}.  

Another popular application of speaker diarization for conversation interaction analysis is the medical doctor-patient interactions. In the system described in \cite{mirheidari2017toward}, the nature of memory problems of a patient is detected from the conversations between neurologists and patients. The speech and language features extracted from the ASR transcripts combined with the speaker diarization results are used to predict the type of disorder. An automated assistant system for medical domain transcription is proposed in \cite{finley2018automated}, which includes the speaker diarization module, ASR module and natural language generation module. The automated assistant module accepts the audio clip and outputs grammatically correct sentences describing the topic of the conversation, subject and subject’s symptom. 

}
{\kh
\subsection{Audio Indexing}

Content-based audio indexing is a well known application domain for speaker diarization. It can provide metainformation such as the content or data type of a given audio data to make information retrieval efficient since search query by machines would be limited by such metadata. The more diverse information was available, the better efficiency could be achieved in retrieving audio contents from a database.

One useful piece of information for the audio indexing would be ASR transcripts to understand the content of speech portions in the audio data. Speaker diarization can augment those transcripts in terms of ``who spoke when'', which was the main purpose of the RT evaluation series \cite{nistrtwebsite}, as discussed in Sections 4.1 and 5.3. The aggregated spoken utterances from speakers by a speaker diarization system also enable per-speaker summary or keyword list-up, which can be used for other query values to retrieve relevant contents from the database. In \cite{guo16remeeting}, we can get a view of how speaker diarization outputs can be linked for information searching in consumer-facing applications. }

{\tp
\subsection{Conversational AI}
Thanks to the advances of ASR technology, the applications of ASR have evolved from simple voice command recognition systems to conversational AI systems. Conversational AI systems, as opposed to voice command recognition systems, have features that are lacking in voice command recognition systems. The fundamental idea of conversational AI is to build a machine that humans can talk to and interact with. In this sense, focusing on an interested speaker in a multiparty setting is one of the most important features of conversational AI. Moreover, speaker diarization becomes an essential feature for conversational AI. For example, conversational AI equipped in a car can pay attention to a specific speaker that is demanding a piece of information from the navigation system by applying speaker diarization along with ASR. 

Smart speakers and voice assistants are the most popular products in which speaker diarization plays a significant role for conversational AI. Since the response time and online processing are the crucial factors in real-life settings, the demand for end-to-end speaker diarization systems integrated into the ASR pipeline is growing. The performance of incremental (online) ASR and speaker diarization of the commercial ASR services are evaluated and compared in \cite{addlesee2020comprehensive}. It is expected that the real-time and low latency aspect of speaker diarization will be more emphasized in the speaker diarization systems in the future since the performance of online diarization and online ASR still have much room for improvement. 

}

\section{Challenges and the Future of Speaker Diarization}
\label{Discussions and Conclusions}
{\nk 

This paper has provided a comprehensive overview of
speaker diarization techniques, highlighting the recent development of deep learning-based diarization approaches.
In the early days, a speaker diarization system was 
developed as a pipeline of sub-modules including front-end processing, 
SAD, segmentation, speaker embedding extraction, 
clustering, and post-processing, leading to a standalone system
without much connection to other components in a given speech application.
With the emergence of the deep learning technology, 
more and more advancements have been made for speaker diarization, 
from a method that replaces a single module into a deep-learning-based method, 
to a fully EEND.
Furthermore,
as the speech recognition technology becomes more accessible,
a trend to tightly integrate 
speaker diarization into the ASR systems has emerged, 
such as benefiting from the ASR output to improve the accuracy of speaker diarization. 
Recently, joint modeling for speaker diarization and speech recognition is investigated in an attempt to 
enhance the overall performance of speaker diarization.
Thanks to these great achievements, speaker diarization systems have already
been used in many applications, including meeting transcription,
conversational interaction analysis, audio indexing, and conversational AI systems.

As we have seen,
tremendous progress has been made for speaker diarization systems. 
Nevertheless, there {\revision is} still much room for improvement.
 As the final remark, 
 we
 conclude this paper by listing the
remaining challenges for speaker diarization
 toward future research and development. 



\paragraph{Online processing of speaker diarization} 

Most speaker diarization methods 
assume that an entire recording can be observed to execute speaker diarization.
However, numerous applications such as meeting transcription systems or smart agents require
very short latency for assigning the speaker.
While there have been several attempts to make online speaker diarization  system
both for clustering-based systems (e.g., \cite{Dimitriadis17}) 
and neural network-based diarization systems (e.g., \cite{zhang2019fully,xue2020online, han2020bw}), 
it still remains as a challenging problem.

\paragraph{Domain mismatch}

A model that is trained on a data in a specific domain often works poorly
on data in another domain.
For example,
it is experimentally known that the EEND model tends to overfit to the
distribution of the speaker overlaps of the training data \cite{fujita2019end}.
Such a domain mismatch issue is universal for any training-based method.
Given the growing interest for trainable speaker diarization systems,
it will become more important to evaluate the ability for handling the variety of inputs.
The international evaluation efforts for speaker diarization such as 
the DIHARD challenge \cite{ryant2018first,ryant2019second,ryant2020third} or VoxSRC \cite{chung2019voxsrc,nagrani2020voxsrc} also have great importance in this direction. 

\paragraph{Speaker overlap}

Overlap of multitalker speech is the inevitable nature of conversation.
For example, an average of 12\% to 15\% of speaker overlap was observed in meeting recordings 
\cite{cetin2006speaker,yoshioka2018recognizing},
and it could even increase in daily conversations \cite{kanda2019guided,barker2018fifth,watanabe2020chime}.
Nevertheless, many traditional speaker diarization systems, especially clustering-based systems,
have only focused on non-overlapping regions and even the overlapping regions are excluded in the evaluation metric \cite{garofolo2006rich}. 
While the topic has been studied for long years 
(e.g. early works \cite{otterson2007efficient,boakye2008overlapped}),
there is a growing interest for handling the speaker overlaps toward better speaker diarization,
including the application of speech separation \cite{xiao2020microsoft},
post-processing \cite{bullock2020overlap,horiguchi2020post}, and 
joint modeling of speech separation and speaker diarization \cite{von2019all,kanda2020joint}. 



\paragraph{Integration with ASR}

Many applications require ASR results along with speaker diarization results.
In the modular combination of speaker diarization and ASR,
some systems locate a speaker diarization system before ASR \cite{medennikov2020stc} while
some systems locate a speaker diarization system after ASR \cite{yoshioka2019advances, park2019speaker}.
Both types of systems showed a strong performance for a specific task,
and determining the best kind of system architecture for the speaker diarization and ASR tasks is still an open problem \cite{raj2020integration}.
Furthermore, there is another line of research to jointly perform 
speaker diarization and ASR \cite{Shafey2019,mao2020speech,kanda2019simultaneous,kanda2020joint}, 
which was introduced in Section \ref{Speaker Diarization in the context of Automatic Speech Recognition}.
The joint modeling approach could leverage the inter-dependency between speaker diarization and ASR
to better perform both tasks.
However, it has not yet been fully investigated whether such joint frameworks perform better than
the well-tuned modular systems.
Overall, the integration of speaker diarization and ASR is one of the hottest topics 
that is still being investigated by many researchers.

\paragraph{Audiovisual modeling} 

Visual information contains a strong clue for the identification of speakers.
For example, the video captured by 
a fish-eye camera was used to improve the accuracy of speaker diarization in a meeting transcription task \cite{yoshioka2019advances}.
The visual information was also used to significantly improve the accuracy of speaker diarization.
for speaker diarization on YouTube video \cite{chung2020spot}.
While these studies showed the effectiveness of visual information,
the audiovisual speaker diarization has yet been rarely investigated compared with
audio-only speaker diarization, and there will be many rooms for improvement.

} 

\bibliographystyle{elsarticle-num-names} 
\bibliography{IEEEfull,conferences,ref}
\end{document}